\documentclass[11pt]{article}
\usepackage{setspace}
\usepackage{graphicx}
\usepackage{lscape}
\usepackage{pdflscape}
\usepackage{afterpage}
\usepackage{pdfpages}
\usepackage{float}
\usepackage{booktabs}
\usepackage{indentfirst}
\usepackage{arydshln}
\usepackage{tikz}
\usepackage{lipsum}
\usepackage{pgffor}
\usepackage{dsfont}
\usepackage{amsfonts}
\usepackage{amsmath}
\allowdisplaybreaks
\usepackage{mathtools}

\usepackage{xfrac}
\usepackage{amssymb}
\usepackage{bigints}
\usepackage{graphicx}
\usepackage{url}				
\usepackage{caption}
\usepackage{subcaption} 
\usepackage{enumitem}
\usepackage{relsize,exscale}
\usepackage{multirow}
\usepackage{natbib}
 
\setcitestyle{authoryear, open={(},close={)}}
\usepackage{datetime}
\newdateformat{monthyeardate}{%
	\monthname[\THEMONTH] \THEYEAR}
\usepackage{titlesec}
\titleformat*{\section}{\large\bfseries}
\titleformat*{\subsection}{\large\bfseries}

\newcounter{parentnumber}
\usepackage{theorem}
\newtheorem{theorem}{Theorem}

\newtheorem{assumption}{Assumption}

\newtheorem{example}{Example}

\newtheorem{proposition}{Proposition}

\newtheorem{remark}{Remark}

\providecommand{\U}[1]{\protect\rule{.1in}{.1in}}
\usepackage{booktabs}
\usepackage{caption}
\begin{document}
	\setstretch{1}
	\title{{\LARGE \textbf{ Partial Identification of Marginal Treatment Effects With Discrete Instruments and Misreported Treatment}\thanks{We would like to thank D\'esir\'e K\'edagni and Otavio Bartalotti for their guidance; Nestor Gandelman, Joydeep Bhattacharya, Kyunghoon Ban, Vitor Possebom,  participants at the events of Sociedad de Economistas del Uruguay (SEU) and Seminars at Iowa State University for their useful comments; and Brent Kreider for sharing the data of the empirical application. .}}}
	\author{ \begin{tabular}{ccc}
	\Large{SANTIAGO ACERENZA}\thanks{ \textit{Universidad ORT Uruguay, Departamento de Economía,
            Blvr. España 2633, 11300 Montevideo, Uruguay (\emph{email:} acerenza@ort.edu.uy)}.} 
	\end{tabular}
	}
	\date{}
	\maketitle
	\newsavebox{\tablebox} \newlength{\tableboxwidth}
\begin{center}
This Draft: \monthyeardate\today
\
		\large{\textbf{Abstract}}
		This paper provides  partial identification results for the marginal treatment effect ($MTE$) when the binary treatment variable is potentially misreported and the instrumental variable is discrete. Identification results are derived under smoothness assumptions. Bounds for both the case of misreported and no misreported treatment are derived. The identification results are illustrated in identifying the marginal treatment effects of food stamps on health.
	\end{center}
	\
		\textbf{Word Count:} 11688.
	\newpage
	\doublespacing
\section{Introduction}\label{Sintro}
\noindent This paper provides partial identification results for the Marginal Treatment Effect ($MTE$) in the presence of measurement error in the treatment variable when only a discrete instrument is available. The discrete instrument case is relevant as many applications in the literature rely on these type of instruments. See for example \cite{AKK}, \cite{AKK2}, \cite{AKK3} and \cite{AKK4}. The discrete nature of the instrument requires identification strategies to recover the $MTE$ that differ from those explored in the previous literature with continuous instruments. \par 
The results of this paper are relevant since it is often true that researchers have access to an instrument with discrete variation (for example, assignment to treatment via an institutional rule), and it is also true that misreporting is a common problem in survey data which is one of the main sources of empirical research.
\par 
In a more general way, our results can serve as a sensitivity analysis tool for when researchers are interested in recovering the $MTE$ in the presence of a discrete instrument and suspect measurement error and have doubts about their parametric assumptions. 
\par 
Researchers mostly work with self-reported data from surveys; such data systematically present reporting problems that lead to measurement error of the treatment status and, consequently, to bias in the treatment effect of interest. The combination of measurement error with discrete instruments has not been explored in the literature, and it is a fairly common situation to encounter. The results in this paper are useful for identifying MTE (which can be used to recover average effects or policy-relevant effects) in the presence of the two previously mentioned problems for identification. 
\par 
In most cases, researchers observe a discrete (often binary) instrument such as assignment to treatment. In these cases, point identification of the $MTE$ (even without measurement error) is not possible, relying only on the standard assumptions of instrument exogeneity and relevance (See for example \cite{BMW}).  In this paper, under a set of restrictions on the severity of measurement error and shape restrictions, we provide partial identification results for the $MTE$ in the presence of measurement error when a discrete instrument is available. 
\par 
The  $MTE$ can help reveal the heterogeneity in the treatment effect. The $MTE$ is relevant in recovering Policy Relevant Treatment Effect parameters ($PRTE$s), Average Treatment Effect ($ATE$), Average Treatment on the Treated ($ATT$), Average Treatment on the Untreated ($ATU$), Local Average Treatment Effects ($LATE$), etc.\footnote{See \cite{HV2}, \cite{HUV}, who show the link between the $MTE$ and those parameters via properly weighting the $MTE$.}
\par 
To achieve partial identification,  we introduce smoothness conditions on the marginal treatment responses ($E[Y_d|V=v]$). To deal with the misreporting of the binary treatment, the analysis relies on treating the unconditional probability of misreporting as given.\footnote{One could alternatively take the results from this paper and assume a known upper bound of this probability and take the union of the bounds derived here.} This can be either interpreted as the researcher having prior knowledge on the possible value of the misclassification rates or as a sensitivity analysis tool where the researcher allows for the possibility of misclassification up to a certain level. Relevance and independence of the instrument is required.   Although partial identification of the $MTE$ will do not imply in general sharp bounds on the $ATE$. It is still a useful tool to move from local effects and generate bounds on an aggregate relevant effect. 
\par 
Empirical research usually combines a measurement error problem with endogeneity and heterogeneity. 
\cite{U} documents in his work, as an example of this, that there is a substantial measurement error in educational attainments in the 1990 U.S. Census. At the
same time, educational attainments are endogenous as treatment variables in return to schooling analyses because, among other possibilities, unobserved individual ability affects both schooling decisions and wages. 
Labor supply response to welfare program participation, in
which the outcome is employment status, and the treatment is welfare program participation is subject to similar issues. Self-reported program participation in survey datasets can be misreported as stated by \cite{HP}. The psychological cost of welfare program participation affects job search behavior and welfare program participation
simultaneously.

\subsection*{Related literature }
\noindent This subsection lists some relevant papers related to the current research paper based on their connections to different aspects of the problem. Namely, misreporting and partial identification of  marginal treatment effects.  
\par 
\textit{Partial identification of $LATE, MTE$ and $ATE$ with endogenous misreported binary treatments and heterogeneous effects}
\par 
\noindent \cite{U} using a binary instrumental variable, derives bounds for $LATE$ with a binary misreported treatment when an instrument is available, and monotonicity of the true (not observed) treatment in the instrument holds. Identification is achieved by exploiting the relationship between the probability of being a complier and the total variation distance\footnote{The total variation distance between two probability measures $P$ and $Q$ on a sigma-algebra $\mathcal {F}$ of subsets of the sample space $\Omega$  is defined via
$\delta (P,Q)=\sup _{A\in {\mathcal {F}}}\left|P(A)-Q(A)\right|$. It can alternatively be defined for probability measures that have densities to be $\frac{1}{2} \int |p-q|d\nu$ where $\nu$ is a measure dominating both probability measures. In the context of \cite{U} paper the total variation distance calculated is $TV_{Y,D}=\frac{1}{2}\int\Big(\sum_{d=0,1}|f_{y,d|Z=1}(y,d)-f_{y,d|Z=0}(y,d)|\Big)d\nu$ where $f_{y,d|z}$ is the joint density of the observed outcome variable and the observed treatment variable conditional on the value of the instrument.} between people assigned to treatment and the ones that are not. The under-identification for $LATE$ is a consequence of
the under-identification for the size of compliers; with no measurement error, one
could compute the size of compliers based on the measured treatment and, therefore,
$LATE$ would be the Wald estimand. The total variation distance plays a key role in determining the sharp identified set in \cite{U}. First, it measures the strength of the instrumental variable; when the total variation distance is positive, the identified set of $LATE$ is a strict subset of the whole parameter space, which implies that $Z$ has some identifying power. Secondly, as shown in \cite{U} lemma 3, the total variation distance is a lower bound for the proportion of compliers which is the under-identified element in the presence of measurement error. \cite{CLT}, \cite{TZ} extend \cite{U}'s results for the case where the instrument can take multiple discrete values. \cite{AK} focuses on bounding the marginal treatment effects when there is a continuous instrument. \cite{KPGJ} using auxiliary information about the possibility of misreporting and under different combinations of the outcome, treatment, and instrumental monotonicity bounds the $ATE$ for a binary outcome.  \cite{VP} focuses on partially identifying the $MTE$ with a continuous instrument and imposing sign and functional relationships between the derivatives of the true propensity score and the observed one with respect to the continuous instrument.   
\par 
This current paper complements the previously mentioned papers. Fundamentally this paper focuses on identifying the $MTE$ when discrete instruments are available. Such a task requires a different set of assumptions than the ones used to recover directly $ATE$, $LATE$, or $MTE$ with continuous instruments. We complement \cite{U}, \cite{KPGJ} and \cite{TZ} because we are interested in identifying $MTE$ (which can then be used to achieve identification of $LATE$ and $ATE$) instead of the $LATE$ and $ATE$. It is also complementing \cite{AK} since their analysis relies on the continuity of the instrument. It is worth noticing that it is more common to observe discrete (mostly binary) instruments such as random selection to receive treatment like in medical studies or random selection to receive a treatment conditional on covariates in social sciences ( e.g., Supplemental Nutrition Assistance Program, SNAP). We complement \cite{VP} since we provide an alternative set of assumptions to identify the $MTE$, and also, we are focusing on a discrete instrument.  
\par 
\textit{Identifying marginal treatment effects with discrete instruments}
\par 
\noindent \cite{BMW}  show how a discrete instrument can be used to identify the marginal
treatment effects under a functional structure that allows for treatment
heterogeneity among individuals with the same observed characteristics and self-selection based on the unobserved gain from treatment. This paper builds upon \cite{BMW} results by considering the case with (endogenous) misreporting and more flexible restrictions (such as shape restrictions instead of parametric assumptions) at the cost of losing point identification. 
The second one is \cite{MST} which using the observed instrumental variables estimates, develops a linear programming approach to recover policy-relevant treatment effects such as the $MTE$. 
This paper differs from it by finding analytical bounds under the different smoothness and shape restrictions. Such bounds permit one to have a first-hand insight into how the assumptions are aiding identification. Estimation of the analytical bounds is simple since it can be performed using their respective sample analogs.
Additionally, \cite{MST} does not allow for the possibility of the treatment to be misreported while here is allowed. In the presence of misreporting, the results from \cite{MST} do not apply directly while the ones derived here do. In the case of no misreporting, our bounds remain valid; in that sense, our results complement the ones from \cite{MST} and \cite{BMW}. 
\subsection*{Outline of the paper}
\noindent The rest of the paper is organized as follows, section \ref{anaF} introduces the main framework and assumptions. Section \ref{Identif} shows the main identification results and illustrates them. 
Section \ref{app} has an application of the identification results to \cite{KPGJ}. Section \ref{con} concludes.  Additional results are collected in the online appendix. Non analytical results on partial identification without additional shape restrictions extending \cite{MST} are included the online appendix section \ref{app1}. Sections \ref{appin} and \ref{appin2} of the appendix focuses on inference for the $ATE$. Section \ref{appchoiceb} discusses how to choose the tuning parameter $b$. Section \ref{appATEBOUND} illustrates the bounds for the $ATE$. Section \ref{appDGP} illustrates the analytical results on a $DGP$. Section \ref{appLM} extends the results with additional monotonicity assumptions. Sections \ref{appDis} and \ref{appDis2}  derives the results for the case when the instrument takes more than $2$ values.\footnote{The case of an instrument taking more than two values can also be seen as the generalization to the case of multiple discrete instruments. This is the case because multiple discrete instruments can be combined in one single multi-valued discrete instrument.} Finally, section \ref{appfig} collects all the figures from the document. 

\section{Analytical Framework}\label{anaF}
\noindent Consider the following framework (\cite{AK}, \cite{HUV} and \cite{HV}):    
\begin{eqnarray}\label{model}
\left\{ \begin{array}{lcl}
     Y &=& Y_1D+Y_0(1-D) \\ 
     D &=& 1\{p(Z)- V\geq 0\}
     \\ 
     D^* &=& D(1-\varepsilon)+(1-D)\varepsilon
     \end{array} \right.
\end{eqnarray}
 Where $Y$ is an outcome variable that can be discrete, continuous, or mixed, the potential outcomes are denoted by $Y_d$, which is the outcome realization for when treatment $D=d$, $D=\{0,1\}$ is a binary unobserved endogenous treatment.  Let $Z\in \mathcal Z=\{z_0,z_1,...,z_k\}$ be a discrete instrument,\footnote{The results will focus on the binary $z$ case but the generalization is natural for more than two values of $z$.} $V$ is a  latent scalar random variable normalized to be uniformly distributed between $(0,1)$.
 $D^*$ is a misreported binary proxy of $D$, the true unobserved treatment status.  $\varepsilon \in \{0,1\}$ is a random variable indicating the presence of misreporting or not. The vector $(Y,D^*,Z)$ is the observed data while $(Y_1, Y_0, D, \varepsilon, V)$ are latent (unobserved).
In the rest of the document, small case letters denote realizations of the respective random variables. 
\par 
\textbf{Object of interest:} In this paper, we care about identifying the $MTE(v^*)$ which is the marginal treatment effect at a particular level $V=v^*$, more precisely, it is defined as $E[Y_1-Y_0|V=v^*]$. 
\par
To identify the $MTE$ in this context, we introduce baseline assumptions that additional assumptions will aid. The baseline assumptions are:
\par 
\begin{assumption}[Random Assignment and Absolute Continuity]\label{Indep11} The following two conditions hold:
\par 
\begin{enumerate}
    \item $Z$ is independent of $(Y_d,V, \varepsilon)$ for all $d=(0,1)$.
\item The distribution of $V$ is absolutely continuous.

\end{enumerate}
\par 
\end{assumption}
The previous assumption and the model structure makes innocuous to say that $V$ is uniform between $[0,1]$ and that $p(Z)=P(D=1|Z)$. 
\begin{assumption}[Relevance]\label{Rele}
Let $Z$ be such that for any $z \in \mathcal{Z}$: 
\begin{enumerate}
    \item $1>p(z)>0$.
\item $p(z)\neq p(z\prime)$ for any $z,z\prime \in  \mathcal{Z}$.
\item  For any $z,z\prime \in  \mathcal{Z}$, we can determine if either $p(z)\leq p(z\prime)$ or $p(z)\geq p(z\prime)$.
\end{enumerate}

\end{assumption}

Assumption \ref{Indep11}-\ref{Rele} include the  instrument independence and validity assumption as in \cite{HV}, \cite{HUV} among others. Assumption \ref{Indep11} requires that $Z$ be a valid instrument, in the sense that it is statistically
independent of the unobservables in the selection equation and the outcome equation. This assumption was used in \cite{AK}. This assumption does not require the measurement
error to be non-differential.\footnote{Non-differential measurement error is that conditional on the unobserved heterogeneity that drives
the selection into treatment, misreporting is independent of the potential outcomes.} Non-differential measurement error combined with assumption \ref{Indep11} implies that misreporting is independent of the outcome conditional on the true treatment, which is in general restrictive. Note that the measurement error can still depend on $z$, but this is through the true treatment since $D^*=D+(1-2D)\varepsilon$. Assumption \ref{Indep11} is restricting the indicator of the existence of measurement error to be independent of $z$ but not the measurement error itself. Assumption \ref{Rele} requires the existence of an instrument that shifts the probability of selection into treatment. In addition, \ref{Rele}   says that, even though the propensity scores cannot be recovered from the
observed data (because $D$ is unobserved in practice), the ascending order of them in $Z$ is still known.
This can be seen as a structural restriction imposed on the true treatment $D$. See \cite{TZ}.\footnote
{An example of sufficient condition  is a constant-coefficient latent-index model.
That is, suppose the treatment is generated by $D = 1(bZ > e)$, where $b$ is a parameter and $e$ is
an error term independent of $Z$. Then, the order of $p(z)$, is
determined by the sign of $b$. It is plausible in many applications that the sign of $b$ can be retrieved
from economic theory. For example, in the study of the returns to schooling, distance to college
is often used as an instrument for completed college education.
In this specific example, the parameter $b$ is negative.}. Under assumption \ref{Rele}, the sign of  $p(z)-p(z\prime)$ for any two $z,z\prime$ is known. 
\par 
Under Assumption \ref{Indep11},we are imposing that 
\begin{eqnarray*}
P(\varepsilon=1)&=&P(\varepsilon=1|Z=z)=P(\varepsilon=1,D=0|Z=z)+P(\varepsilon=1,D=1|Z=z)
\\ &=& P(\varepsilon=1|D=0, Z=z)P(D=0|Z=z)+P(\varepsilon=1|D=1,Z=z)P(D=1|Z=z)
\\ &=& P(\varepsilon=1|V>P(z), Z=z)P(D=0|Z=z)+P(\varepsilon=1|V \leq P(z),Z=z)P(D=1|Z=z)
\end{eqnarray*}
With the false-positive probability $P(\varepsilon=1|D=0, Z=z)$ and the false-negative probability
$P(\varepsilon=1|D=1, Z=z)$. 
\par 
We introduce the working example that will help interpret the assumptions and results through the rest of the document. 
\begin{example}
The researcher is interested
in measuring marginal returns of  recieving the Supplemental Nutrition Assistance Program (SNAP) on food security. It is well documented that underreporting of SNAP exists. In this case, the variable $Y$ is a binary outcome of being food secure, and $D$ is the true indicator
for being a SNAP recipient. The variable $Z$ is the indicator of having certain assets in the household or having cars exempt from an asset test that recipients have to complete (see \cite{KPGJ} and references therein). 
\par 
The latent variable $V$ could
be interpreted as the stigma cost of SNAP as in \cite{Welfare2}. As stated by \cite{Welfare1} stigma is acknowledged as one of the determinants of welfare participation, and there is wide evidence that it negatively
affects take-up rates.
\par 
Let $Y_1$ be the potential food security status for someone
on SNAP, and $Y_0$ when the same individual does not receive it. $Y_d$ can be correlated with the stigma cost $V$. As noted by \cite{Food1}, internalized stigma may lead to food insecurity if it causes or intensifies isolation from social support systems that would allow access to food. Additionally, as stated by \cite{Food2} stigma manifestations lead to food inequities through a series of mediating mechanisms experienced and enacted by targets of the stigma that
undermine healthy food consumption, contribute to
food insecurity, and ultimately impact diet quality. In that sense, psycho-social processes represent how individuals respond to stigma, which ultimately shapes their food selection, purchasing, and consumption behaviors. Enacted and anticipated
stigma are characterized as significant stressors, and individuals may cope with these stressors
through unhealthy eating behaviors or irrational choices that increase the likelihood of food insecurity. This is then implicitly saying that stigma could be correlated with the potential outcomes. 
\par 
The variable $D^*$
is the individual’s reported (observed) indicator for SNAP recipiency. In this context, the last part of assumption \ref{Indep11} is consistent with saying that the stigma cost is also determining the misreporting behavior of the individual says $\varepsilon= 1\{f(V)\geq e\}$, if the function of the stigma cost is big enough to pass some threshold $e$ the individual chooses to misreport consistent with \cite{HP}. Additionally, the assumption is consistent with random misreporting; one could think that individuals make errors when answering the survey question about SNAP recipiency with no intention. In such case $\varepsilon= 1\{f(\eta)\geq 0\}$ where $\eta$ is independent of $V, Y_d$. 
\end{example}
\begin{remark}
Imposing that the instrument is independent from the misclassification decision may not be appropriate in many empirical contexts although we claim it is valid here. To illustrate when it is not valid using the current example for instance, if in the SNAP example, the instrument $Z$ is a result of the political forces that regulate SNAP implementation in each state the assumption would not hold. These political forces may influence how people perceive the benefits and costs associated with welfare participation. If those perceived costs are associated with individual willingness to lie about SNAP participation, then $Z$ is not independent of the decision to misreport, implying that the assumption does not hold in this empirical example. 
\end{remark}
\par
Besides the previously mentioned baseline assumptions, the following assumption is introduced. 
\begin{assumption}[Smoothness]\label{Lip}
There exists known constants, $b$ such that for any pairs $v_1\neq v_2$ in the support of $V$:
\begin{align}
  \begin{array}{lcl}    
-b|v_1-v_2|\leq  E[Y_d|V=v_1]-E[Y_d|V=v_2]\leq b|v_1-v_2|
 \end{array}
\end{align} 
\end{assumption}
\cite{KKKL} introduces smoothness conditions for $E[Y_d|D=d]$ to bound the $ATE$ without an instrument and treatment exogeneity; this approach has the same spirit. In this case, we can build on their insight to provide bounds for the $MTE$ using similar smoothness conditions. 
The previous assumption states the degree of smoothness of the marginal treatment responses ($E[Y_d|V=v]$). Generally speaking, we may interpret our identification analysis in this section as a conditional one
indexed by $b$. Furthermore, we may conduct a sensitivity analysis by looking at different
values of $b$. The parameter $b$ is the Lipschitz constant which serves as a measure of smoothness. In this case we are assuming a maximum level of smoothness $b$. 
\par
Assumption \ref{Lip} is restricting the functional form for the marginal responses, but considering all possible functionals in the lipschitz family with smoothness parameter $b$ or smaller instead of a particular parametric family (like for example linear functions). It is stating the degree of smoothness of the potential responses without assuming a particular functional form of it. In this sense $E[Y_d|V=v]$ could be for example linear $E[Y_d|V=v]=\mu_d+a_dv$ (in which case $b=a_d$) or quadratic  $E[Y_d|V=v]=\mu_d+a_dv+c_dv^2$ (in which case $b=a_d+2c_d$) among different possibilities. This assumption introduces constraints in the underlying selection mechanism since is imposing restrictions on how the potential outcomes behave in relationship to the underlying cost of selecting into treatment. The smoothness assumption also relies on the choice of the Lipschitz constant which makes the result sensitive to the choice. This later point is discussed in the online appendix.   Assumption \ref{Lip} might more appropriately be called something like bounded slope, bounded rate of change, or Lipschitz continuity of the $MTR$ functions but they are directly impacting the degree of parsimony of the functions, so we call it smoothness.\footnote{It is worth noticing that in
the standard analysis on the $MTE$, the normalization of $V$ does not change any content of the model, but it does change the interpretation of the Lipschitz condition. This is because it is not the same to impose a Lipschitz condition on the conditional mean of $Y_d$ on $E$ where $E$ has a normal distribution, than to put it on $V=F_E(E)$ which is uniform.}
\par 
The following remark adapted from \cite{KKKL} is relevant to understand what this type of assumption is imposing on the marginal treatment responses. 
\begin{remark}
An alternative way of bounding the rate of change in the marginal treatment responses is to impose further global restrictions in addition to monotonicity such as concavity. The approach used in this paper  imposes
restrictions directly on the rate of change in its nature, whereas the combination of
concavity and monotonicity restricts the rate of change indirectly. There is no clear dominance between each of these ways of imposing restrictions except the belief the researcher has on the behaviour of the marginal treatment responses. 
\end{remark}
\par
More generally one could say $b^{\prime}|v_1-v_2|\leq  E[Y_d|V=v_1]-E[Y_d|V=v_2]\leq b|v_1-v_2|$ as stated by \cite{KKKL}, furthermore, letting $b^{\prime}=0$ and saying  $E[Y_d|V=v_1]-E[Y_d|V=v_2]\leq b|v_1-v_2|$ for $v_1>v_2$ would be combining monotonicity of the treatment responses with assumption \ref{Lip}. More precisely:
\begin{assumption}\label{lip2}
There exists known constants, $b_1,b_0>0$ such that for any pairs $v_1\geq v_2$ in the support of $v$:
\begin{align}
  \begin{array}{lcl}    
0\leq  E[Y_1|V=v_1]-E[Y_1|V=v_2]\leq b_1(v_1-v_2)
 \end{array}
\end{align}    
\begin{align}
  \begin{array}{lcl}    
0\leq  E[Y_0|V=v_1]-E[Y_0|V=v_2]\leq b_0(v_1-v_2)
 \end{array}
\end{align}             
Or more generally: 
\begin{align}
  \begin{array}{lcl}    
0\leq  E[Y_d|V=v_1]-E[Y_d|V=v_2]\leq b(v_1-v_2)
 \end{array}
\end{align} 
Where $b=\max\{b_0,b_1\}$
\end{assumption}

\begin{example}[Continued]

In the context of SNAP, a binary treatment, and food security, a binary outcome, one could model the relationship using a bivariate probit model. Nevertheless, this can be restrictive since it implies a known joint distribution of the unobservables and a parametric index structure. Alternatively, one could choose to allow for all the models with $b \leq 0.5$.  This is consistent with the bivariate probit models and allows for more generality by relaxing the normality assumption.
\end{example}

\subsection*{Identification breakdown}
\noindent Note that following  \cite{HUV} and their standard assumptions (\ref{Indep11}-\ref{Rele} above), without further restrictions the $MTE$ at the level of heterogeneity $v^*$ ($E[Y_1-Y_0|V=v^*]$) is not identified in this setting with discrete instruments and a misreported treatment.
\par
From standard results, we get:
\begin{eqnarray*}
E[Y|Z=z]&=&\int_0^{p(z)}E[Y_1|V=v]dv+\int_{p(z)}^{1}E[Y_0|V=v]dv
\\
E[Y|Z=z]-E[Y|Z=z\prime]&=&\int_{p(z\prime)}^{p(z)}E[Y_1|V=v]-E[Y_0|V=v]dv
\end{eqnarray*}
The second equation takes the difference of the first equation for any two values of the instrument connecting the observed shift in $Y$ caused by changes in $z$ and the underlying treatment effect for all the individuals affected by such a change of the instrument.
\par 
For any given $z$, say $z\prime$ we can get:  
\begin{align}\label{E2}
  \begin{array}{lcl}
     P[D^*=1|Z=z\prime]&=&P[D=1|D^*=D,Z=z\prime]P[D^*=D|Z=z\prime] \\&+&P[D^*=1|D^*\neq D,Z=z\prime]P[D^*\neq D|Z=z\prime].
 \end{array}
\end{align}
Where the first equality is because we are conditioning on $D^*=D, D^* \neq D$ and applying the properties of probabilities. This last equation reflects that the observed propensity score for the proxy of the true treatment variable conditional on $z\prime$ equals the share of treated individuals who at that particular $z\prime$  report treatment status correctly multiplied by the probability of reporting correctly, plus the share of not treated individuals who at that particular $z\prime$  report treatment status incorrectly multiplied by the probability of reporting incorrectly. 
\par 
The previous expressions depend on unobserved components. While $ E[Y|Z=z\prime], P[D^*=1|Z=z\prime]$ are observed,   $P[D^*=D|Z=z\prime],P[D=1|D^*=D,Z=z\prime]$ and $p(Z)$  are not, which without further assumptions do not allow for identification of the true propensity score and also of the $MTE$. 
\par 
If $p(z)$ was observed and $z$ was continuous, then the $MTE(v^*)$ would be identified as $\frac{\partial E[Y|P(Z)=v^*] }{\partial v^*}=\frac{\int_0^{v^*}E[Y_1|V=v]dv
    + \int_{v^*}^1E[Y_0|V=v]}{\partial v^*}=E[Y_1|V=v^*]-E[Y_0|V=v^*]$. So this displays the two main identification challenges, the non-continuity of $z$ and the fact that $p(z)$ is not observed.

\par

Before proceeding to the identification results, it is worth showing the main elements of the current work and how they differentiate from previous identification results of $MTE$ with discrete instruments. It is also relevant to show the role of misreporting.  
\par 
From the observed data if there is no misreporting from assumptions \ref{Indep11}-\ref{Rele} one can identify:
\begin{eqnarray*}
E[YD|Z=z]&=&E[Y_1 1\{p(Z)- V\geq 0\}|Z=z]=E[Y_1 |p(z)- V\geq 0,Z=z]p(z)  \\
&=&E[Y_1 |p(z)- V\geq 0]p(z)=\int_0^{p(z)}E[Y_1|V=v]dv
\end{eqnarray*}
The first equality comes from the definition of the model, the second one from the laws of probability, the third one by the independence of $Z$ from $Y_1, V$ and the last one from the properties of conditional expectations and the normalization that $V$ is marginally uniform.  
\par 
Similarly, we have: 
\begin{eqnarray*}
E[Y(1-D)|Z=z]&=&\int_{p(z)}^{1}E[Y_0|V=v]dv
\end{eqnarray*}
This then implies the equality expressed at the beginning of this subsection: 
\begin{eqnarray}\label{Insight}
E[Y|Z=z]&=&\int_0^{p(z)}E[Y_1|V=v]dv+\int_{p(z)}^{1}E[Y_0|V=v]dv
\end{eqnarray}

Without misreporting the propensity score $p(z)$ is identified and, given assumptions \ref{Indep11}-\ref{Rele} index sufficiency holds and thus $E[YD|Z=z]=E[YD|p(Z)=p]$. So we can rewrite the previous equalities as functions of $p$ instead of $z$. Where $p \equiv p(z)$.
\par 
In this context without differentiability of $p$ the key insight from \cite{BMW} is to introduce parametric restrictions  that for example say that $E[Y_d|V=v]=\mu_d
+a_d v$ and thus $E[Y_1-Y_0|V=v]=\mu+a v$, where $\mu\equiv \mu_1-\mu_0$ and $a \equiv a_1-a_0$. Additionally define $c=\mu_0+\frac{a_0}{2}$.  In this case
\begin{eqnarray*}
E[YD|p(Z)=p]&=& \mu_1p+\frac{a_1}{2}p^2
\\
E[Y(1-D)|p(Z)=p]&=& \mu_0(1-p)+\frac{a_0}{2}(1-p^2)
\\
E[Y|p(Z)=p]&=&c+\mu p+\frac{a}{2}p^2
\end{eqnarray*}
Then from $E[YD|p(Z)=p]$ for different values of $p$ (at least two which is enough with a binary instrument) we can solve for $\mu_1,a_1$. Similarly for $a_0,\mu_0$ from $E[Y(1-D)|p(Z)=p]$. 
\par
Note that then given the marginal treatment responses, $E[Y_d|V=v]$ is identified, so it is the $MTE$ as their difference. 
\par 
One might not be willing to assume particular parametric specifications for the conditional expectations of the potential outcomes since they are restrictive. One of the contributions of the current work is relaxing such restrictions and still recovering analytically tractable expression for the bounds of the $MTE(v^*)$.
\par 
The current work relates \cite{MST} in the following way. \cite{MST}  relies on recovering the set of marginal treatment responses consistent with observed $IV$-like estimands. In this setting, such strategy would rely on finding all the candidates $E[Y_d|V=v]$ functions consistent with:
\begin{eqnarray*}
\frac{E[Y|Z=z]-E[Y|Z=z\prime]}{p(z)-p(z\prime)}&=& \int_{p(z\prime)}^{p(z)}\frac{E[Y_1|V=v]}{p(z)-p(z\prime)}dv+\int_{p(z\prime)}^{p(z)}\frac{-E[Y_0|V=v]}{p(z)-p(z\prime)}dv
\\
&=& \int_{0}^{1}\frac{E[Y_1|V=v]}{p(z)-p(z\prime)}1\{v \in (p(z\prime), p(z))\}dv\\
&+&\int_{0}^{1}\frac{-E[Y_0|V=v]}{p(z)-p(z\prime)}1\{v \in (p(z\prime), p(z))\}dv 
\\
&\equiv& \int_{0}^{1}E[Y_1|V=v]w_1dv+\int_{0}^{1}E[Y_0|V=v]w_0dv 
\end{eqnarray*}
\par 
Their strategy relies on the fact that $w_1 w_0$ are known (or identified). In the case of misreporting, where we do not know exactly the rate of false positives and false negatives for every value of $z$, we have that $p(z)$, and thus, the weights are not identified. This makes the current work to differ from the existing literature since developed computational methods rely on the weights being known or identified. In this context one of the main contributions of the current paper is working in the context where the weights are not identified but actually can be partially identified.  
\par 
In section \ref{Identif} we start from the same insight as \cite{MST}, but instead of solving a linear problem, we aid identification with shape restrictions to get analytical bounds on the $MTE$. In the online appendix an extension of \cite{MST} is discussed without aiding identification with shape restrictions by solving the same linear problem as in  \cite{MST} but for different values of $p(z)$ score in the identified set. As the identified set of  $p(z)$ is not finite, the solution can only be approximated. 
\section{Identification results }\label{Identif}
\noindent In subsection \ref{Identifnomis} identification without misreporting will be discussed. Subsection \ref{3.2} incorporates misreporting. 
\subsection*{Identification without misreporting}\label{Identifnomis}
\noindent Note than since 
\begin{eqnarray*}
E[Y|Z=z] &=& \int_0^{p(z)}E[Y_1|V=v]dv + \int_{p(z)}^1E[Y_0|V=v]dv.
\end{eqnarray*}
For any values $z$ and $z\prime$ with $p(z) \geq p(z\prime)$:
\begin{eqnarray*}
  E[Y|Z=z]-E[Y|Z=z\prime] &=& \int_{p(z\prime)}^{p(z)}E[Y_1|V=v]-E[Y_0|V=v]dv\\
  &=&\int_{p(z\prime)}^{p(z)}\bigg(E[Y_1|V=v]-E[Y_1|V=v^*]-E[Y_0|V=v]
  \\&+&E[Y_0|V=v^*]+E[Y_1|V=v^*]-E[Y_0|V=v^*]\bigg)dv 
  \\ &\leq&  \int_{p(z\prime)}^{p(z)}2b|v-v^*|dv+\int_{p(z\prime)}^{p(z)}MTE(v^*)dv
    \\
    &=&  \int_{p(z\prime)}^{p(z)}2b|v-v^*|dv+MTE(v^*)(p(z)-p(z\prime)).
\end{eqnarray*}
Where we are adding and subtracting the marginal treatment responses ($E[Y_d|V=v^*]$) related to the marginal treatment effect  at the $v^*$ of interest ($E[Y_1|V=v^*]-E[Y_0|V=v^*]$), then using $E[Y_d|V=v]-E[Y_d|V=v^*]\leq b|v-v^*|$ twice and also the definition of $MTE$. Similarly we can get:
\begin{eqnarray*}
  E[Y|Z=z]-E[Y|Z=z\prime] &\geq&  \int_{p(z\prime)}^{p(z)}-2b|v-v^*|dv+MTE(v^*)(p(z)-p(z\prime)).
\end{eqnarray*}
Then:
\footnotesize
\begin{eqnarray*}
\frac{E[Y|Z=z]-E[Y|Z=z\prime]-\int_{p(z\prime)}^{p(z)}2b|v-v^*|dv}{p(z)-p(z\prime)} \leq MTE(v^*) \leq \frac{E[Y|Z=z]-E[Y|Z=z\prime]+\int_{p(z\prime)}^{p(z)}2b|v-v^*|dv}{p(z)-p(z\prime)}.
\end{eqnarray*}
\normalsize
The bounds depend on the propensity score and the difference of the propensity score for different values of $z$. 
\begin{remark}
Note that if one integrates the bounds for the $MTE$ one does not point identify $LATE$. In particular, integrating $v^*$ over $p(z\prime)$ and $p(z)$ yields the following bounds for $LATE$:  
\footnotesize
\begin{eqnarray*}
\frac{E[Y|Z=z]-E[Y|Z=z\prime]-\frac{2b}{3}(p(z)-p(z\prime))^3}{p(z)-p(z\prime)} \leq LATE(p(z\prime),p(z)) \leq \frac{E[Y|Z=z]-E[Y|Z=z\prime]+\frac{2b}{3}(p(z)-p(z\prime))^3}{p(z)-p(z\prime)}.
\end{eqnarray*}
\normalsize
This is because the way the bounds are derived, a quantity that is bigger (or smaller) of the numerator of $LATE$ is central to derive the bounds for the $MTE$. The method to derive bounds for the $MTE$ is not exactly an extrapolation of $LATE$ since there is no unique way to do that given assumption  \ref{Lip}.  More precisely, the way the bounds are computed, assumption  \ref{Lip} is applied at every point $v^*$ without consideration of the joint restrictions for pairs of evaluation points such as $v^*,v^{**}$. Additionally, implications of \ref{Lip} are not necessarily fully exploited in the constructive identification approach. In this sense, the bounds are neither functional nor point-wise sharp. 
\end{remark}
\begin{remark}
Then the partial identification analysis of the $ATE$ starts from the
well-known Manski’s worst-case bound.
This formulation of the identification region reveals that the identification power becomes weak when the upper and lower bounds for $Y_d$ are large. An advantage of the proposed method is that it does not require the existence of upper and lower bounds (although it requires a tuning parameter $b$). In this case as shown in the online appendix, for example, the $ATE$ can be bounded above by: 
\footnotesize
\begin{eqnarray*}
ATE&\leq&\frac{E[Y|Z=z]-E[Y|Z=z\prime]}{p(z)-p(z\prime)}+\frac{2b}{3}\frac{p(z)^3-p(z\prime)^3}{p(z)-p(z\prime)}-b\frac{p(z)^2-p(z\prime)^2}{p(z)-p(z\prime)}+b
\end{eqnarray*}
\normalsize
If $Y_d$ is bounded, then the previous bound on the $ATE$ is complemented with the Manski worst case bounds: 
\footnotesize
\begin{eqnarray*}
ATE&\leq&\min\{ Y_{u1}-Y_{l0} ,\frac{E[Y|Z=z]-E[Y|Z=z\prime]}{p(z)-p(z\prime)}+\frac{2b}{3}\frac{p(z)^3-p(z\prime)^3}{p(z)-p(z\prime)}-b\frac{p(z)^2-p(z\prime)^2}{p(z)-p(z\prime)}+b \}
\end{eqnarray*}
\normalsize
Note there is a $b$ such that the proposed bounds are numerically the same as the worst case bounds in the case the outcome variable is bounded. 
\end{remark}
\subsection*{Identification of the MTE with misreporting}\label{3.2}
\noindent In order to identify the $MTE$ first we need to identify $p(z)$. In \cite{AK} such identification is discussed. Subsequent subsections, builds upon the results from \cite{AK}. \par
For clarity, let $\Delta_p \equiv p(z)-p(z'),  \Delta_{D^*Z}(z',z) \equiv P(D^*=1|Z=z)-P(D^*=1|Z=z'), \alpha \equiv P(\varepsilon=1) $, then let the bounds derived in \cite{AK} be: 
\begin{eqnarray*}
\Delta_{pl} &\equiv&  \Delta_{D^*Z}(z',z), \\
\Delta_{pu} &\equiv& \min\left\{1,2\alpha+\Delta_{D^*Z}(z',z), 2(1-\alpha)-\Delta_{D^*Z}(z',z)\right\},  \\
p_l(z) &\equiv& \max\left\{ P(D^*=1\vert Z=z)-\alpha, \alpha- P(D^*=1\vert Z=z)\right\}, \\
p_u(z) &\equiv& \min\left\{ P(D^*=1\vert Z=z)+\alpha, (1-\alpha)+ P(D^*=0\vert Z=z)\right\}.
\end{eqnarray*}
The bounds on the propensity score rely on any given level of unconditional misreporting ($\alpha$). 
 Misreporting enters in two ways. First of all, if the probabilities of misreporting are unknown, $p(z)$ is no longer identified (see \cite{AK} for more details), which then creates a problem since such quantity appears systematically in the bounding strategies. To solve this problem, we use the previously defined bounds on the misreporting probabilities. 
\par 
The lack of point identification of $p(z)$ affects the strategy using smoothness restrictions. See for example that: 
\begin{eqnarray*}
  E[Y|Z=z]-E[Y|Z=z\prime]&\leq&  \int_{p_l(z\prime)}^{p_u(z)}2b|v-v^*|dv+MTE(v^*)(p(z)-p(z\prime))
\end{eqnarray*}
The bound of $MTE$ depends on the sign of it since it is not always true that  $MTE(v^*)(p(z)-p(z\prime)) \leq MTE(v^*)(p_u(z)-p_l(z\prime))$. These considerations are taken into account in theorem \ref{TL}. 
\par 
\textit{Smoothness assumptions for marginal treatment responses}\label{3.2.2}
\par 

\noindent The  bounds from theorem \ref{TL} builds on the following inequalities due to assumption \ref{Lip}
\begin{eqnarray}\label{lipi1}
  E[Y|Z=z]-E[Y|Z=z\prime]&\leq&  \int_{p_l(z\prime)}^{p_u(z)}2b|v-v^*|dv+MTE(v^*)(p(z)-p(z\prime))
\end{eqnarray}
\begin{eqnarray}\label{lipi2}
  E[Y|Z=z]-E[Y|Z=z\prime]&\geq&  -\int_{p_l(z\prime)}^{p_u(z)}2b|v-v^*|dv+MTE(v^*)(p(z)-p(z\prime))
\end{eqnarray}
Where the  inequalities use the smoothness assumption as in the previous section without misreporting and fact that $\int_{p(z\prime)}^{p(z)}2b|v-v^*|dv \leq \int_{p_l(z\prime)}^{p_u(z)}2b|v-v^*|dv$. Note that we have a  sufficient condition for identifying the sign of the $MTE(v^*)$. 
If $E[Y|Z=z]-E[Y|Z=z\prime]-  \int_{p_l(z\prime)}^{p_u(z)}2b|v-v^*|dv\geq 0$ then the $MTE(v^*)$ is positive. If $E[Y|Z=z]-E[Y|Z=z\prime]+  \int_{p_l(z\prime)}^{p_u(z)}2b|v-v^*|dv\leq 0$ then the $MTE(v^*)$ is negative. 
\par 
If $E[Y|Z=z]-E[Y|Z=z\prime]-  \int_{p_l(z\prime)}^{p_u(z)}2b|v-v^*|dv\geq 0$ then from equations \ref{lipi1} and \ref{lipi2} combined with the bounds on $p(z)-p(z\prime)$ we get: 
\begin{eqnarray*}
MTE^+(v^*)_{lb}=\frac{E[Y|Z=z]-E[Y|Z=z\prime]-  \int_{p_l(z\prime)}^{p_u(z)}2b|v-v^*|dv}{\Delta_{pu}} \\
MTE^+(v^*)_{ub}=\frac{E[Y|Z=z]-E[Y|Z=z\prime]+  \int_{p_l(z\prime)}^{p_u(z)}2b|v-v^*|dv}{\Delta_{pl}}
\end{eqnarray*}
Note that the form of the bounds depend on $|v-v^*|$. In the integral of the absolute value is where the relative position of $v^*$ with respect of $p_l(z\prime),p_u(z)$ will matter. Note that if the $v^*$ of interest is such that $v^*\leq p_l(z\prime)$, the integral involving $|v-v^*|$ is  $ [\frac{p_u(z)^2-p_l(z\prime)^2}{2}+v^*(p_l(z\prime)-p_u(z))]$. If the $v^*$ of interest is such that $v^*\geq p_u(z)$ then $[\frac{-p_u(z)^2+p_l(z\prime)^2}{2}+v^*(-p_l(z\prime)+p_u(z))]$. If the $v^*$ of interest is  such that $p_l(z\prime)\leq v^*\leq p_u(z)$ then $[v^{*2}-v^*(p_u(z)+p_l(z\prime))+(\frac{p_u(z)^2+p_l(z\prime)^2}{2})]$

The following theorem summarizes the previous discussion. 

\begin{theorem}\label{TL}
If assumptions \ref{Indep11}-\ref{Rele} and \ref{Lip} holds. Then the following bounds are valid: 
\begin{enumerate}
   
\item If $E[Y|Z=z]-E[Y|Z=z\prime]-  \int_{p_l(z\prime)}^{p_u(z)}2b|v-v^*|dv\geq 0$:
\begin{eqnarray*}
MTE^+(v^*)_{lb}=\frac{E[Y|Z=z]-E[Y|Z=z\prime]-  \int_{p_l(z\prime)}^{p_u(z)}2b|v-v^*|dv}{\Delta_{pu}} \\
MTE^+(v^*)_{ub}=\frac{E[Y|Z=z]-E[Y|Z=z\prime]+  \int_{p_l(z\prime)}^{p_u(z)}2b|v-v^*|dv}{\Delta_{pl}}
\end{eqnarray*}
\item If $E[Y|Z=z]-E[Y|Z=z\prime]+  \int_{p_l(z\prime)}^{p_u(z)}2b|v-v^*|dv\leq 0$:
\begin{eqnarray*}
MTE^-(v^*)_{lb}=\frac{E[Y|Z=z]-E[Y|Z=z\prime]-  \int_{p_l(z\prime)}^{p_u(z)}2b|v-v^*|dv}{\Delta_{pl}} \\
MTE^-(v^*)_{ub}=\frac{E[Y|Z=z]-E[Y|Z=z\prime]+  \int_{p_l(z\prime)}^{p_u(z)}2b|v-v^*|dv}{\Delta_{pu}}
\end{eqnarray*}
\item Otherwise: 
\begin{eqnarray*}
MTE(v^*)_{lb}=\max\{MTE^-(v^*)_{lb},MTE^+(v^*)_{lb}\} \\
MTE(v^*)_{ub}=\min\{MTE^-(v^*)_{ub},MTE^+(v^*)_{ub}\}
\end{eqnarray*}
\end{enumerate}
\end{theorem}
\begin{remark}
In some situations like in the case of SNAP, one could be willing to assume that for every level of heterogeneity $v$, $P(Y_1<Y_0|V=v)=1$ holds which means that receiving SNAP is not making anyone more food insecure. This is the ``treatment cannot hurt'' assumption or known as the monotone treatment response assumption in the partial identification literature. In such a case, we would be imposing the sign of the $MTE$  even if we cannot extract it from $E[Y|Z=z]-E[Y|Z=z\prime]-  \int_{p_l(z\prime)}^{p_u(z)}2b|v-v^*|dv\geq 0$ or $E[Y|Z=z]-E[Y|Z=z\prime]+  \int_{p_l(z\prime)}^{p_u(z)}2b|v-v^*|dv\leq 0$
\end{remark}
\begin{remark}
The previous bounds are not assuming there is a known support for $Y$ if the nature of $Y$ is bounded then the previous bounds change in the following way:
\begin{eqnarray*}
\Tilde{MTE}(v^*)_{lb}&=&\max\{MTE(v^*)_{lb},Y_{1l}-Y_{0u}\} \\
\Tilde{MTE}(v^*)_{ub}&=&\min\{MTE(v^*)_{ub},Y_{1u}-Y_{0u}\}
\end{eqnarray*}
\end{remark}
The previous theorem is extended for discrete instruments taking more than 2 values in the online appendix. 
\par 
A researcher might be interested in combining the monotonicity assumption on the treatment responses and the smoothness assumption. So instead of using assumption \ref{Lip}, the researcher might be willing to use \ref{lip2}. This result is collected in the online appendix.  
\par
The choice of $b$ is not arbitrary. The fact that it operates as a tuning parameter might lead to a discretionary use of it to get the desired result; different ways of choosing this parameter are discussed in the online appendix.  
\par 
The previous identification results are illustrated in the appendix. 
\section{Application: $MTE$ of SNAP  on child health  when participation is endogenous and misreported}\label{app}
\noindent In this section, the developed methods are applied to get bounds on the $MTE$. We then integrate them over $v^*$ to get bounds on the $ATE$  of receiving SNAP on the outcome of being food insecure. As stated by \cite{KPGJ} SNAP, formerly known as the Food Stamp Program, is
by far the largest food assistance program in the United States and, as such, constitutes a crucial
component of the social safety net in the United States. In any given month during 2009, SNAP assisted more than 15 million children, and it is estimated that
nearly one in two American children will receive assistance during their childhood. Concluding about the program's impact is complex due to two of the fundamental problems studied in this paper.  First, a selection problem arises because the decision to participate in SNAP is unlikely to be exogenous. On the contrary, unobserved factors such as expected future health status, parents’ human capital characteristics,
financial stability, and attitudes towards work and family are all thought to be jointly related to
participation in the program and health outcomes such as food security. Families may decide to participate precisely because they expect to be food insecure or in poor health.
Second, a nonrandom measurement error problem arises because a large fraction of food stamp recipients fails to correctly report their program participation in household surveys. Using administrative data matched with data from the Survey of Income and Program Participation (SIPP), for example, \cite{BD} find that errors in self-reported receipt of food stamps exceed 12 percent and are related to respondents’ characteristics, including their true participation status, health outcomes, and demographic attributes. \cite{MMS} provide evidence of extensive underreporting of food stamps in the SIPP, the Current Population Survey (CPS), and the Panel Study of Income Dynamics (PSID). 
\par 
In this context  \cite{KPGJ} studies the average effects using the December Supplement of the 2003  Current Population Survey (CPS). In the data, we can observe a self-reported measure of food stamp receipt over the past year, food insecurity over the past year, and the ratio of income to the poverty line.\footnote{For further details about the data see \cite{GK}. In there, they state that just over 40 percent of the households report receiving food stamps, and the food insecurity rate among self-reported recipients is 17.9 percentage points higher than among eligible non-recipients (52.3 percent vs. 34.4).}
\par 
As \cite{KPGJ} states, the data is rich enough to allow the construction of  instrumental variables for SNAP participation used in previous literature. In particular, state identifiers in the CPS apply a more traditional instrumental variable (IV) assumption based on cross-state variation in program eligibility rules.\footnote{In general terms, program eligibility rules are income requirements (most households must meet both gross and net income
limits to qualify for SNAP benefits), resource requirements (households must also meet a resource limit in their bank accounts), work requirements (If you are an able-bodied adult without dependents, between the ages of 18 and 49, and able to work but currently unemployed, you may only be eligible
for SNAP benefits for three months within a three-year
period) and other eligibility requirements (to be eligible for SNAP benefits, households must
also, meet other conditions in addition to the income
and resource requirements, such as everyone in
your household having, or have applied for, a social
security number). To establish the income and resource requirements, each state computes asset tests, but there is variation in how these states evaluate the assets of individuals. More specifically, they may or may not include certain assets that will affect the individuals' eligibility.} Merging the Urban  Institute’s database of state program rules with the CPS data   \cite{KPGJ}  create two instrumental variables: an indicator for whether the state uses a simplified semi-annual reporting requirement for earnings and an indicator for whether cars are exempt from the asset test.\footnote{For more details on the construction see \cite{KPGJ}.} These instrumental variables if valid and independent allows using the current methods to bound the $MTE$. For example, if one is willing to assume that the state variation in the asset test is exogenous, then instrument independence is satisfied. It is worth noticing that only $LATE$ can be identified from an instrumental variable regression under individual heterogeneity. In this sense, the methods developed here can be used to recover bounds on the average treatment effect, and complement \cite{KPGJ} results. In this context, the $MTE$ at a particular level of $v^*$ represents the treatment effect receiving SNAP has for a particular level of stigma. Stigma mat is connected with the potential outcomes since, as stated by \cite{Food2} stigma manifestations affect health outcomes (and food security as such).  
\par 
More precisely, in this context to illustrate our methods, $Y$ is a food insecurity indicator over the past year, $D^*$ is the self-reported SNAP participation (subject to potential measurement error as stated before), $Z$ is a binary variable for cars exempt from the asset test.  
\par 
Concerning choosing $b$ for the sake of exposition, we present the results here for several potential values of $b$. In the online appendix  we discuss different methods on how to choose $b$ which could be used in this context. Intuitively choosing $b$ is restricting the degree of smoothness the $MTE$ would have. One can draw a parallelism between choosing a linear form (a low $b)$ for the $MTE$ versus choosing a high order polynomial (high $b$). The linear form is rather restrictive on the behavior of higher-order derivatives (and thus smoothness) compared with the polynomial. A researcher choosing $b$ for SNAP should consider what he thinks the underlying decision-maker optimization problem looks like. If the expected utility, for example, has a quadratic form, or the optimal expected demand of food security is linear under both receiving and not receiving SNAP, then the expected benefit on the optimal choice of food security both under getting SNAP and not getting it could be considered linear.
\par 
Consistently with \cite{KPGJ} a treatment cannot hurt assumption ($P(Y_1<Y_0|V=v)=1$) will be introduced. Making then the conservative upper bounds of the $MTE$ to be $0$. 
\par 
The following table summarizes the data and shows the average and median characteristics in the sample. 
\begin{table}[!htbp] 
\centering 
\caption{\textit{Summary statistics}}
\begin{tabular}{l|ccc}
& \multicolumn{3}{c}{}   \\ 
 &\textit{Mean} &\textit{Standard Deviation}  &\textit{Median}\\  \hline\hline\\
Food insecure ($Y$)  &0.42&0.49&0  \\ 
Reporting being on SNAP ($D$)  &0.41&0.49&0    \\   
Cars exempt from the asset test ($Z$)  &0.30&0.46&0    \\ 
Income to poverty line ratio  &0.75&0.36&0.75    \\ 
$N$ & 2707&-&-    \\\hline
\end{tabular}
\end{table}
We can see that around forty-two percent of the people are food insecure, while forty-one report receiving SNAP. Thirty percent have their cars excluded from the asset test, which implies that thirty percent of the individuals in the sample live in states where cars are excluded from the asset test. On average (and in the median), individuals in this sample are below the poverty line. 
\par 
Each of the graphs in the following figure is computed in the following way. For any given level of $b$ and $\alpha$, point estimates of the bounds are constructed for the
$MTE$ at different values $v^*$ using their sample analogs. To decide which type of bound to use, the relative position of the sample analog estimates of the bounds for the propensity score is calculated. The maximum level of $\alpha$ is twenty percent which is chosen as an arbitrary big upper bound above the existing results from \cite{KPGJ}. 
\par 
See Figure \ref{F1}.

In orange, we have the upper and lower bounds assuming $\alpha=0.2$. In blue, we have the upper and lower bounds assuming $\alpha=0.1$, and in red, we report the upper and lower bounds assuming $\alpha=0$ (no misreporting). The limits of the $Y$-axis are the worst-case upper bound ($0$ under treatment cannot hurt) and the worst-case lower bound ($-1$). The $X$-axis goes from $0$ to $1$, the different potential values of $v^*$. We can see that the bounds have identification power over different regions of $v^*$ support. We can see that when the level of misreporting decreases, the bounds become tighter since there is less lack of identification due to misreporting. Similarly, when $b$ decreases, we also get tighter bounds consistent with reducing the potential functional forms of the marginal treatment responses. 
\par 
These bounds are computed by fixing a level of $\alpha$. If, for example, in the case of $b=0.1$, the researcher is not interested in $\alpha=0.2$ rather in $\alpha \leq 0.2$ then all the region between the upper orange curve and the lower orange curve is the identified set consistent with $b=0.1$ and all the $\alpha$'s less or equal to twenty percent. 
\par 
The length of the identified set becomes tighter in the region between the estimated observed propensity scores ($\widehat{P}[D^*=1|Z=1]=0.49,\widehat{P}[D^*=1|Z=0]=0.38$) since there is more information being used to bound the $MTE$.   
\par
The previous display also implies a simple way of computing bounds of a parameter of interest such as the $ATE$. It is known that $ATE=\int_{0}^1MTE(v^*)dv^*$. Then we know that $\int_{0}^1MTE_{lb}(v^*)dv^* \leq ATE \leq \int_{0}^1MTE_{ub}(v^*)dv^*$. So then we can approximate bounds for the $ATE$ as: 
\begin{eqnarray*}
\frac{1}{N_{v^*}}\sum_{v^*}\widehat{MTE}_{lb}(v^*) &\equiv& \widehat{ATE}_{lb}  \\
\frac{1}{N_{v^*}}\sum_{v^*}\widehat{MTE}_{ub}(v^*) &\equiv& \widehat{ATE}_{ub}
\end{eqnarray*}
Where $N_{v^*}$ is the number of grid points where $MTE(v^*)$ was evaluated and where $\widehat{MTE}$ are the bounds from the previous graphs. On the online appendix estimates of the bounds on the $ATE$ are computed. 
In the online appendix, there is also an alternative way of estimating (and doing inference) on the $ATE$ for the case of smoothness restrictions. On the online appendix a method for asymptotic normality for an outer-set of the $ATE$ in the case of no misreporting and with smoothness conditions is developed. On the online appendix a method for asymptotic normality for an outer-set of the $ATE$ in the case of misreporting, treatment cannot hurt assumption, and smoothness conditions can be found. 
\par 
The  $ATE$ bounds are easily estimated and used for inference, since as shown in the appendix, each component can be replaced by their sample analogs, which themselves are asymptotically normal, and thus, by the continuous mapping theorem the upper bounds and lower bounds for the $ATE$ are also. Then, an asymptotically valid bootstrap procedure can be used to build
confidence intervals for the entire identified set, such as those constructed by \cite{MN}. The population identification region 
is an interval $[L, U]$, we can estimate each side of the interval with consistent and asymptotically normal estimators $\widehat{L}, \widehat{U}$ and via this procedure we get  confidence interval $[\widehat{L}-z_{\frac{\alpha+1}{2}}\widehat{\sigma}_l/\sqrt{N}, \widehat{U}+z_{\frac{\alpha+1}{2}}\widehat{\sigma}_u/\sqrt{N}] \equiv [\widehat{L}_{\alpha}, \widehat{U}_{\alpha}]$ such that
\begin{eqnarray*}\label{Cutoff}
\lim_{N \rightarrow \infty}P([L, U] \in [\widehat{L}_{\alpha}, \widehat{U}_{\alpha}])=1-\alpha 
\end{eqnarray*}
\par 
In order to control for covariates $X$ such that independence of $Z$ holds conditional on it in a tractable manner, we can assume $E[Y_d|V=v,X=x]=m_d(v)+\beta X$. Then,
\begin{eqnarray*}
E[Y|Z=z,X=x] &=& \int_0^{p(z,x)}E[Y_1|V=v,X=x]dv + \int_{p(z,x)}^1E[Y_0|V=v,X=x]dv
\\ &=&  \int_0^{p(z,x)}m_1(v)dv + \int_{p(z,x)}^1m_0(v)dv+\beta X
\end{eqnarray*}
Where $P(z,x)=P(D=1|X=x,Z=z)$. We can then any two values of $Z$: 
\begin{eqnarray*}
  E[Y|Z=z,X=x]-E[Y|Z=z\prime,X=x]
  &=&\int_{p(z\prime,x)}^{p(z,x)}[m_1(v)-m_0(v)]dv 
\end{eqnarray*}
We can then follow a similar display is in Section \ref{Identif}. In this case, we can estimate $E[Y|Z=z,X=x]$ with a partial linear regression while $p(z,x)$ is estimated with a non-parametric regression. 
\par 
So far, we have reported results for the instrument taking only two values. The data set used in this problem counts with two potential instruments. The already used one, and an eligibility criterion specifying if earners report twice a year or not. Based on this, a three-valued instrument can be constructed related to the intensity of the likelihood of receiving SNAP. That is, it takes $0$ if both instruments take the value $0$, takes the value $1$ if either of them takes the value $1$, and it takes the value $2$ if both of them take the value $1$. The details of how the bounds look for more discrete non-binary instruments and the particular case of an instrument taking three values are collected in the online appendix.   In such a case, the length of the identified set for the $ATE$ becomes smaller; this is intuitive since we now have a more exogenous variation to exploit. We we illustrate it with the case of $b=0.5, \alpha=0.1$. Additionally, we can see that the form of the identified set of the $MTE$ changes, since now the regions rely on the different exogenous variation in zones where previously, only the shape restrictions could be used. These results are collected in the online appendix.    
\section{Conclusions}\label{con}
\noindent In this paper, we provided partial identification results for the Marginal Treatment Effect in the presence of measurement error and a discrete instrument building over \cite{MST}, \cite{BMW} and \cite{AK}. To do so,  given the discrete nature of the instruments, we introduced smoothness restrictions.   
\par 
Results are illustrated via a numerical example and quantifying the marginal treatment effect of SNAP on food insecurity, a case in which measurement error and endogeneity of treatment are known to be an issue.    
\par 
In a more general way, our results can serve as a sensitivity analysis tool for when researchers are interested in recovering the $MTE$ in the presence of a discrete instrument and suspect measurement error, and have doubts about their parametric assumptions. This sensitivity analysis is executed by varying $\alpha, b$.  
\par
If no measurement error exists, the results from this paper provided analytical partial identification results of the $MTE$ in the presence of discrete  instruments that can serve as a complement for the already existing results. 
\singlespace
  
\pagebreak
\newpage
\pagebreak
\setcounter{table}{0}
\renewcommand\thetable{A.\arabic{table}}
\setcounter{figure}{0}
\renewcommand\thefigure{A.\arabic{figure}}
\setcounter{equation}{0}
\renewcommand\theequation{A.\arabic{equation}}
\appendix
\begin{center}
	\huge
	Supporting Information

	(Online Appendix)
\end{center}
\doublespacing
\normalsize
\setcounter{table}{0}
\renewcommand\thetable{A.\arabic{table}}
\setcounter{figure}{0}
\renewcommand\thefigure{A.\arabic{figure}}
\setcounter{equation}{0}
\renewcommand\theequation{A.\arabic{equation}}
\setcounter{theorem}{0}
\renewcommand\thetheorem{A.\arabic{theorem}}
\section{Identification without shape restrictions}\label{app1}
\noindent The object of interest as noted is $E[Y_1-Y_0|V=v^*]$. Following \cite{MST}  this can be expressed as:
\begin{eqnarray}\label{MTEs}
E[Y_1|V=v^*]-E[Y_0|V=v^*]=\int_{0}^{1}E[Y_1|V=v]w_{1}dv+ \int_{0}^{1}E[Y_0|V=v]w_{0}dv
\end{eqnarray}
 Where $w_1=\delta(v^*)$, $w_0=-\delta(v^*)$ and $\delta(v^*)$ is Dirac delta measure assigning all the mass at $V=v^*$. 
\par 
We can express $E[Y|Z=z]-E[Y|Z=z\prime]$ our $IV$-like estimand as:
\begin{eqnarray}\label{IV-like}
E[Y|Z=z]-E[Y|Z=z\prime]&=& \int_{p(z\prime)}^{p(z)}E[Y_1|V=v]dv-\int_{p(z\prime)}^{p(z)}\-E[Y_0|V=v]dv
\\ \nonumber
&=& \int_{0}^{1}E[Y_1|V=v]1\{v \in (p(z\prime), p(z))\}dv\\ \nonumber
&+&\int_{0}^{1}-E[Y_0|V=v]1\{v \in (p(z\prime), p(z))\}dv 
\\ \nonumber
&\equiv& \int_{0}^{1}E[Y_1|V=v]\omega_1dv+\int_{0}^{1}E[Y_0|V=v]\omega_0dv  \nonumber
\end{eqnarray}
Where $\omega_1=1\{v \in (p(z\prime), p(z))\}$ and $\omega_0=-1\{v \in (p(z\prime), p(z))\}$. 
\par 
Note that $1\{v \in (p(z\prime), p(z))\}$ is not known since $p(z)$ is not known. Nevertheless, for any fixed $\{p(z),p(z\prime)\}$, for $E[Y_d|V=v] \in \mathcal{M}$, $d=1,0$ and $\mathcal{M}$ being a convex space, we known from \cite{MST} that since the equations \ref{MTEs} and \ref{IV-like} define linear operators convexity is carried onto the space of solutions of equation  \ref{MTEs} subject to \ref{IV-like}. Then this allows to define a linear programming as in \cite{MST} and take into account  the implementation considerations they make to get upper bounds and lower bound for the $MTE$ by solving respectively: 
\begin{eqnarray}
\max (\min)_{E[Y_1|V=v],E[Y_0|V=v] \in \mathcal{M}}&&\int_{0}^{1}E[Y_1|V=v]w_{1}dv+ \int_{0}^{1}E[Y_0|V=v]w_{0}dv
\\ \nonumber
\text{Subject to}
\\ 
E[Y|Z=z]-E[Y|Z=z\prime]&=&\int_{0}^{1}E[Y_1|V=v]\omega_1(p)dv+\int_{0}^{1}E[Y_0|V=v]\omega_0(p)dv  \nonumber
\end{eqnarray}
Which then give as a solution an interval defined as $MTE_{lb}(v^*,p), MTE_{ub}(v^*,p)$. 
Where $\omega_d(p), MTE_{lb}(.,p), MTE_{ub}(.,p)$ is stating the dependence of the program to a particular $p(z), p(z\prime)$. 
\par 
The following procedure can be repeated for every $p(z), p(z\prime) \in \mathcal{P}$, the identification region of the propensity score ( $\mathcal{P}$ defined in \cite{AK})  and then  the set of possible values for the $MTE$ is $\bigcup_{p \in \mathcal{P}}\big(MTE_{lb}(v^*,p), MTE_{ub}(v^*,p)\big)$. In practice the calculation cannot be made for every $p$ since it is infinite-dimensional, but the solution can be approximated taking several grid points in the space $\mathcal{P}$.  

\section{Inference for the $ATE$ with no misreporting and smoothness conditions}\label{appin}
\noindent Let $\Delta_Y \equiv E[Y|Z=z]-E[Y|Z=z\prime]$. Let $p_1\equiv p(z)$ and $p_0 \equiv p(z\prime)$. We can use the upper bounds on the $MTE$ to construct upper bounds on the $ATE$ (a similar display would apply for the lower bounds). 
\par 
Note that from the bounds developed under smoothness conditions it is true that the following are upper bounds for the $MTE(v^*)$ (namely $MTE(v^*)_{ub}$): 
\begin{eqnarray}\label{Eqab1}
MTE(v^*)_{ub1} \equiv \frac{\Delta_y+2b[\frac{p_1^2-p_0^2}{2}+v^*(p_0-p_1)]}{p_1-p_0}  \quad \text{if} \quad  v^*< p_0 
\end{eqnarray}
\begin{eqnarray}\label{Eqab2}
MTE(v^*)_{ub2} \equiv \frac{\Delta_y+2b[v^{*2}+\frac{p_1^2+p_0^2}{2}-v^*(p_0+p_1)]}{p_1-p_0}  \quad \text{if} \quad p_0< v^*< p_1 
\end{eqnarray}
\begin{eqnarray}\label{Eqab3}
MTE(v^*)_{ub3} \equiv \frac{\Delta_y+2b[\frac{-p_1^2+p_0^2}{2}+v^*(-p_0+p_1)]}{p_1-p_0}  \quad \text{if} \quad  p_1< v^* 
\end{eqnarray}
Then from the fact that $ATE=\int_0^1MTE(v^*)dv^*$ we can see that: 
\begin{eqnarray}\label{Eqab4}
ATE_{ub}&=&\int_0^1MTE(v^*)_{ub}\\ 
&=& \int_0^{p_0}MTE(v^*)_{ub1}+ \int_{p_0}^{p_1}MTE(v^*)_{ub2}+\int_{p_1}^1 MTE(v^*)_{ub3} \nonumber
\end{eqnarray}
Which then combining equations \ref{Eqab1}-\ref{Eqab4} and after some calculus we get: 
\begin{eqnarray}\label{Eqab5}
ATE_{ub}&=&\frac{\Delta_y}{p_1-p_0}+\frac{2b}{3}\frac{p_1^3-p_0^3}{p_1-p_0}-b\frac{p_1^2-p_0^2}{p_1-p_0}+b
\end{eqnarray}
Which is a smooth continuous function of $p_1,p_0,\Delta_y$ (except at $p_1=p_0$ which is ruled out by assumption) then if the estimators of $p_1,p_0,\Delta_y$ are asymptotically normal (which is the case under standard conditions since they are sample analogs) we get by the continuous mapping theorem that the estimator of $ATE_{ub}$ is also asymptotically normal. Then we can perform valid asymptotic inference on the bounds on the $ATE$. The bound is an outer set because, as pointed out in the main document, if the variables $Y_1, Y_0$ are naturally bounded then, so it is the $ATE$, the bounds here do not incorporate that aspect.

\section{Inference for the $ATE$ with  smoothness conditions and treatment cannot hurt assumption}\label{appin2}
\noindent As in section \ref{appin}, let $\Delta_Y \equiv E[Y|Z=z]-E[Y|Z=z\prime]$. Now as there is misreporting let  $p_{1u}\equiv p_u(z), p_{1l} \equiv p_l(z), p_{0u} \equiv p_u(z\prime)$ and $p_{0l} \equiv p_l(z\prime)$.
\par 
Also let the lower bound of the difference of the probabilities  as in the main text to be $\Delta_{pl}$. In this case adding the assumption that $Y_1-Y_0\leq 0$ we are imposing an upper bound on the $MTE$ and $ATE$ to be $0$. We are also imposing information on the sign of it which leads to the following lower bounds for the $MTE(v^*)$

\begin{eqnarray}\label{Eqac1}
MTE(v^*)_{lb1} \equiv \frac{\Delta_y-2b[\frac{p_{1u}^2-p_{0l}^2}{2}+v^*(p_{0l}-p_{1u})]}{\Delta_{pl}}  \quad \text{if} \quad  v^*< p_{0l} 
\end{eqnarray}
\begin{eqnarray}\label{Eqac2}
MTE(v^*)_{lb2} \equiv \frac{\Delta_y-2b[v^{*2}+\frac{p_{1u}^2+p_{0l}^2}{2}-v^*(p_{0l}+p_{1u})]}{\Delta_{pl}}  \quad \text{if} \quad p_{0l}< v^*< p_{1u} 
\end{eqnarray}
\begin{eqnarray}\label{Eqac3}
MTE(v^*)_{lb3} \equiv \frac{\Delta_y-2b[\frac{-p_{1u}^2+p_{0l}^2}{2}+v^*(-p_{0l}+p_{1u})]}{\Delta_{pl}}  \quad \text{if} \quad  p_{1u}< v^* 
\end{eqnarray}
Then from a similar display as in section \ref{appin} we get: 
\begin{eqnarray}\label{Eqac5}
ATE_{lb}&=&\frac{\Delta_y-\frac{2b}{3}(p_{1u}^3-p_{0l}^3)+b(p_{1u}^2-p_{0l}^2)-b(p_{1u}-p_{0l})}{\Delta_{pl}}
\end{eqnarray}

We know that $\Delta_y, \Delta_{pl}$ can be estimated with the sample analogs, and they are well-behaved estimators that, under standard central limit theory, are asymptotically normal. Note that from the main text  $p_{1u}\equiv \min\left\{ P(D^*=1\vert Z=z)+\alpha, (1-\alpha)+ P(D^*=0\vert Z=z)\right\} , \\ p_{0l} \equiv \max\left\{ P(D^*=1\vert Z=z\prime)-\alpha,  \alpha- P(D^*=1\vert Z=z\prime)\right\}$ where the $\max,\min$ operators make the asymptotic normality of their sample analogs not possible. But note that  $p_{1u}\leq P(D^*=1\vert Z=z)+\alpha$ and $p_{0l} \geq P(D^*=1\vert Z=z\prime)-\alpha$. So if the researcher is willing to assume he is using levels of $\alpha$ that are such that $P(D^*=1\vert Z=z)+\alpha \leq 1$ and $P(D^*=1\vert Z=z\prime)-\alpha \geq 0$ and also theirs sample analogs, then, he can use $P(D^*=1\vert Z=z)+\alpha, P(D^*=1\vert Z=z\prime)-\alpha$ instead of $p_{1u},p_{0l}$ as the bounds on the probabilities. In such a case, the sample analogs of these outer bounds are asymptotically normal by the usual central limit theory.\footnote{Note that in the development of the bounds on the $MTE$ with misreporting the particular form of the bounds for $p(z)$ was never used. In that sense, the previous results still hold just that now we change tighter bounds of $p(z)$ for wider ones.} 
In this case then, since the bound on the $ATE$ is a smooth continuous function of $p_{1u},p_{0l},\Delta_y, \Delta_{pl}$ and since the sample analogs of  $P(D^*=1\vert Z=z)+\alpha,P(D^*=1\vert Z=z\prime)-\alpha,\Delta_y, \Delta_{pl}$ are asymptotically normal, by standard results we get by the continuous mapping theorem that the estimator of $ATE_{lb}$ is also asymptotically normal. Then we can perform valid asymptotic inference on the bounds on the $ATE$. 
This bound is an outer set because, as pointed out in Section \ref{appin} and also because we are not using the tightest possible bounds on $p(z)$.

\section{The choice of $b$}\label{appchoiceb}
\noindent In some applications,  choosing $b$ involves some subjective belief about the maximum size of treatment effects or, as above, on the underlying behavior of unobservable taste parameters. Identification results are
obtained conditional on those beliefs. One possible route to choose $b$ as proposed by \cite{KKKL} formally is to rely on
Bayesian inference using pre-samples or information from prior elicitation. Using existing experimental results or previous research, one may obtain a posterior distribution regarding $b$ and use a high quantile of the posterior distribution as a possible value
of $b$. 
\par 
An alternative way of choosing $b$ in the current paper and the application to SNAP is if there are previous results on $LATE$ or $ATE$ for SNAP, we could choose $b$ to be such that is consistent with previous studies on the topic. 
\par 
Yet another way would be in the same spirit of \cite{AK2} and a-priory decide on the bigger  (or worst case) class of functions the researcher is willing to accept as potential marginal treatment responses. In that sense, if the researcher is willing, for example, to accept the idea that all functions between $0$ and $1$ with $b=2$ or less are candidates, then he should present the report for all the values of $b$ consistent with this notion.  
\par 
The previous ideas all rely on the researcher either having a belief ex-ante or auxiliary data. An alternative way of choosing the $b$ from inside the given data itself is the following.  Suppose the support of the instrument has at least three values, $z_0,z_1,z_2$ and respective propensity scores $p_0,p_1,p_2$. If the researcher was willing to use only information on $z_0,z_2$ to get bounds on the $MTE$ at different values of $v^*$ between $0$ and $1$ then notice the following equality: 
\begin{eqnarray*}
 E[Y|Z=z_2]-E[Y|Z=z_1] &=& \int_{p_1}^{p_2}MTE(v)dv\\
\end{eqnarray*}
The obtained bounds on the $MTE$ could be plugged in to get: 
\begin{eqnarray*}
 E[Y|Z=z_2]-E[Y|Z=z_1] &\leq& \int_{p_1}^{p_2}MTE_{ub}(v)dv\\
 E[Y|Z=z_2]-E[Y|Z=z_1] &\geq& \int_{p_1}^{p_2}MTE_{lb}(v)dv\\
\end{eqnarray*}
Similarly 
\begin{eqnarray*}
 E[Y|Z=z_1]-E[Y|Z=z_0] &\leq& \int_{p_0}^{p_1}MTE_{ub}(v)dv\\
 E[Y|Z=z_1]-E[Y|Z=z_0] &\geq& \int_{p_0}^{p_1}MTE_{lb}(v)dv\\
\end{eqnarray*}
Then the range of $b$ could be chosen as consistent with the previous set of inequalities supported by the data. This strategy would require the instrument to take at least three values and would also imply not using all the information available to get the tightest possible bounds on the $MTE$ conditional on $b$. Nevertheless, this would bring a way to discipline the value of $b$ to be consistent with the observed data. 
\par 
A similar logic can be applied with the $ATE$. Take for example the bounds for the $ATE$ derived above in the case of no misreporting.  
\footnotesize
\begin{eqnarray*}
ATE&\leq&\frac{E[Y|Z=z]-E[Y|Z=z\prime]}{p(z)-p(z\prime)}+\frac{2b}{3}\frac{p(z)^3-p(z\prime)^3}{p(z)-p(z\prime)}-b\frac{p(z)^2-p(z\prime)^2}{p(z)-p(z\prime)}+b
\end{eqnarray*}
\normalsize
Then note that, 
\begin{eqnarray*}
\frac{ATE-\frac{E[Y|Z=z]-E[Y|Z=z\prime]}{p(z)-p(z\prime)}}{\frac{2}{3}\frac{p(z)^3-p(z\prime)^3}{p(z)-p(z\prime)}-\frac{p(z)^2-p(z\prime)^2}{p(z)-p(z\prime)}+1}&\leq&b 
\end{eqnarray*}
Now note that, from our data we have estimates for $ E[Y|Z]$ and $p(z)$. Then plugging in these estimates we get:  
\begin{eqnarray*}
\frac{ATE+0.272727273}{0.510466667 }&\leq&b 
\end{eqnarray*}
Then say for example, we recover from \cite{RM} that $ATE=-0.162$, then we get $b=0.22$. 
\par 
Note that this method also suggest an alternative way to choose $b$ in the spirit of \cite{MP}. In this sense, we can choose the maximum level of $b$ to be the one such that there is no average treatment effect. Thus,
\begin{eqnarray*}
b&=& \frac{\frac{E[Y|Z=z]-E[Y|Z=z\prime]}{p(z)-p(z\prime)}}{\frac{2}{3}\frac{p(z)^3-p(z\prime)^3}{p(z)-p(z\prime)}-\frac{p(z)^2-p(z\prime)^2}{p(z)-p(z\prime)}+1}
\end{eqnarray*}
Which would be in the example $b=0.534270483$

\section{Bounds on the $ATE$}\label{appATEBOUND}

\begin{table}[ht]
\caption{Bounds on the $ATE$}
\centering 
\begin{tabular}{|c|c|c|c|c|c|}
\hline
\multicolumn{2}{|c|}{\begin{tabular}[c]{@{}c@{}}$\alpha=0.2$, \\ $b=1$\end{tabular}}        & \multicolumn{2}{c|}{\begin{tabular}[c]{@{}c@{}}$\alpha=0.2$, \\ $b=0.5$\end{tabular}}      & \multicolumn{2}{c|}{\begin{tabular}[c]{@{}c@{}}$\alpha=0.2$, \\ $b=0.1$\end{tabular}}    \\ \hline
LB                                           & UB                                           & LB                                           & UB                                          & LB                                          & UB                                         \\ \hline
-1.00                                        & 0.00                                         & -0.96                                        & 0.00                                        & -0.48                                       & -0.04                                      \\ \hline
\multicolumn{2}{|c|}{\begin{tabular}[c]{@{}c@{}}$\alpha=0.1$, \\ $b=1$\end{tabular}}        & \multicolumn{2}{c|}{\begin{tabular}[c]{@{}c@{}}$\alpha=0.1$, \\ $b=0.5$\end{tabular}}      & \multicolumn{2}{c|}{\begin{tabular}[c]{@{}c@{}}$\alpha=0.1$, \\ $b=0.1$\end{tabular}}    \\ \hline
LB                                           & UB                                           & LB                                           & UB                                          & LB                                          & UB                                         \\ \hline
-0.94                                        & 0.00                                         & -0.81                                        & 0.00                                        & -0.38                                       & -0.09                                      \\ \hline
\multicolumn{2}{|c|}{\begin{tabular}[c]{@{}c@{}}$\alpha=0$, \\ $b=1$\end{tabular}}          & \multicolumn{2}{c|}{\begin{tabular}[c]{@{}c@{}}$\alpha=0$, \\ $b=0.5$\end{tabular}}        & \multicolumn{2}{c|}{\begin{tabular}[c]{@{}c@{}}$\alpha=0$, \\ $b=0.1$\end{tabular}}      \\ \hline
LB                                           & UB                                           & LB                                           & UB                                          & LB                                          & UB                                         \\ \hline
-0.72                                        & -0.02                                        & -0.50                                        & -0.05                                       & -0.28                                       & -0.18                                      \\ \hline
\end{tabular}
\label{ATEB}
\end{table}
 \newpage
\section{Details of the $DGP$}\label{appDGP}

\noindent To illustrate the results, we build the following  DGP motivated by the application to SNAP. As stated by \cite{KPGJ} the case of SNAP is sensitive to misreporting. It is more likely to observe people receiving SNAP and erroneously say they are not receiving it, than people not receiving it saying that they do. This is consistent with setting $(1-D)\varepsilon=0$, which means no one misreports receiving when they do not receive it. 
Consider the following $DGP$:
\begin{eqnarray}\label{DGP2}
\left\{ \begin{array}{lcl}
     Y &=& DV +(1-D)\frac{V}{4}  \\
     D &=& 1\{Z- V\geq 0\}
     \\
    D^* &=&  D(1-\varepsilon)
    \\  
    \varepsilon &=& 1\{V\leq 0.15 \}
     \end{array} \right.
\end{eqnarray}
$V\sim U(0,1)$, $Z$ takes values  be $0.7$ or $0.1$ with probability  $1/2$, $Z$ is independent of $V$. Note $|E[Y_1|V=v_1]-E[Y_1|V=v_2]|=|v_1-v_2|$, $|E[Y_0|V=v_1]-E[Y_0|V=v_2]|=\frac{1}{4}|v_1-v_2|$ so $b$ can be set to $1$. Note $MTE(v)=\frac{3}{4}v$. The marginal treatment responses $E[Y_1|V=v]=v,E[Y_0|V=v]=\frac{1}{4}v$ are monotonic in $v$. $\alpha=P(\varepsilon=1)=0.15$. 
\par
\begin{figure}[ht]
\centering
\caption{Results from the DGP}\label{F2}
\includegraphics[width=\textwidth]{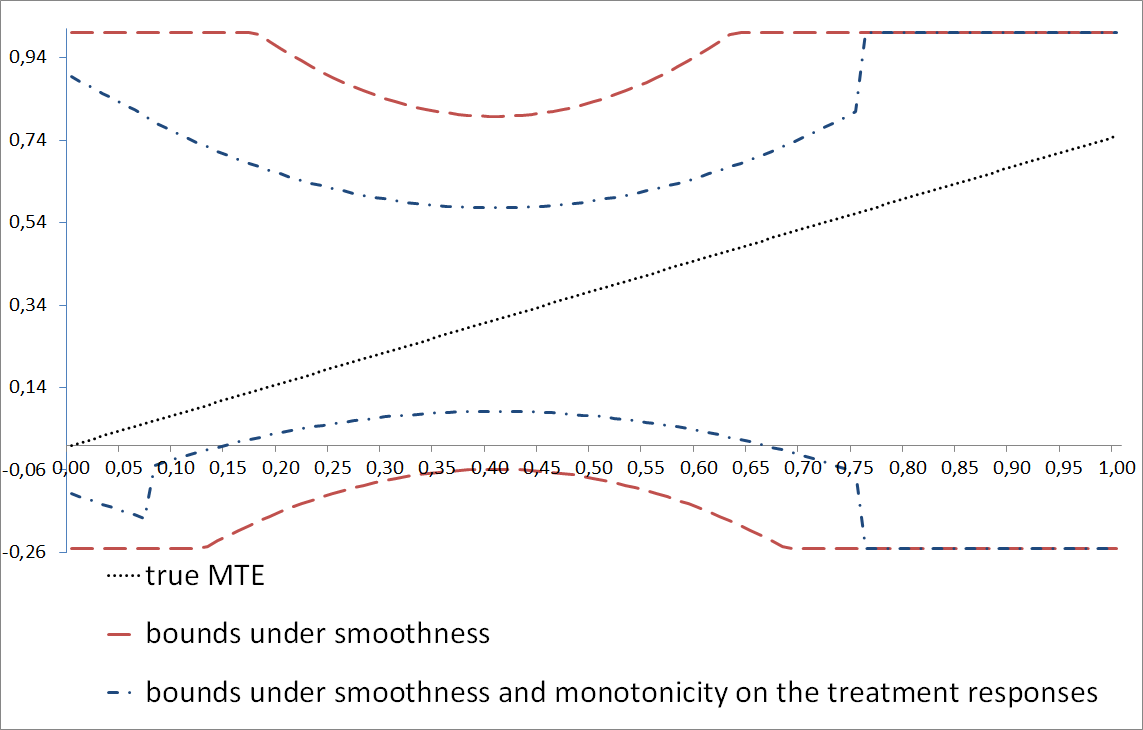}
\end{figure}

 
See Figure \ref{F2}. 
\par 
The limit of the $Y$ axis is the worst case upper bounds ($1$) and lower bounds ($-0.25$). The black dotted line is the true $MTE$ curve. The red lines represent the bounds from using the smoothness assumption alone. We can see that even though it improves from the worst-case bounds in this particular DGP, it cannot recover the sign of the $MTE$. Finally, the blue lines represent the combination of monotonicity of the treatment responses and smoothness. In such cases improvements over only smoothness are achieved, and the sign of the $MTE$ is recovered in certain regions. 
In general, we can see that the bounds improve over the worst-case bounds and have identifying power on the $MTE$.

\section{Identification with monotonicity assumption  on the treatment responses and  smoothness}\label{appLM}
This appendix collects the result for the case when the researcher is willing to assume both monotonoicity and smoothness of the treatment responses.   In such a case note then that for some $v^*$ between $p(z),p(z\prime)$: 
\footnotesize
\begin{eqnarray*}
  E[Y|Z=z]-E[Y|Z=z\prime] &=& \int_{p(z\prime)}^{p(z)}E[Y_1|V=v]-E[Y_0|V=v]dv\\
  &=&\int_{p(z\prime)}^{p(z)}\bigg(E[Y_1|V=v]-E[Y_1|V=v^*]-E[Y_0|V=v]
  \\&+&E[Y_0|V=v^*]+E[Y_1|V=v^*]-E[Y_0|V=v^*]\bigg)dv 
  \\ &=& [p(z)-p(z\prime)]MTE(v^*) \\
  &+&\int_{p(z\prime)}^{p(z)}E[Y_1|V=v]-E[Y_1|V=v^*]dv + \int_{p(z\prime)}^{p(z)}E[Y_0|V=v^*]-E[Y_0|V=v]dv \\
  &=& [p(z)-p(z\prime)]MTE(v^*) \\
  &+&\int_{p(z\prime)}^{v^*}E[Y_1|V=v]-E[Y_1|V=v^*]dv +\int_{v^*}^{p(z)}E[Y_1|V=v]-E[Y_1|V=v^*]dv\\
  &+& \int_{p(z\prime)}^{v^*}E[Y_0|V=v^*]-E[Y_0|V=v]dv+\int_{v^*}^{p(z)}E[Y_0|V=v^*]-E[Y_0|V=v]dv \\
\end{eqnarray*}
\normalsize
Note that between $p(z\prime), v^*$, every $v$ is smaller than $v^*$. Then by assumption \ref{lip2}  for $v^*$ bigger than $v$ we have  $0 \leq E[Y_1|V=v^*]-E[Y_1|V=v] \leq b(v^*-v) $,  then $0 \geq -E[Y_1|V=v^*]+E[Y_1|V=v] \geq b(v^*-v)$ thus between $p(z\prime)$ and $v^*$, $\int_{p(z\prime)}^{v^*}E[Y_1|V=v]-E[Y_1|V=v^*]dv\leq 0$. Similarly between $v^*$ and $p(z)$ we get $\int_{v^*}^{p(z)}E[Y_1|V=v]-E[Y_1|V=v^*]dv\leq \int_{v^*}^{p(z)}b(v-v^*)dv$.  Also $\int_{p(z\prime)}^{v^*}E[Y_0|V=v^*]-E[Y_0|V=v]dv\leq \int_{p(z\prime)}^{v^*}b(v^*-v)dv$, $\int_{v^*}^{p(z)}E[Y_0|V=v^*]-E[Y_0|V=v]dv \leq 0$. Then: 
\footnotesize
\begin{eqnarray*}
  E[Y|Z=z]-E[Y|Z=z\prime] &\leq& 
  [p(z)-p(z\prime)]MTE(v^*) \\
  &+&\ \int_{v^*}^{p(z)}b(v-v^*)dv\\
  &+& \int_{p(z\prime)}^{v^*}b(v^*-v)dv \\
  &=& [p(z)-p(z\prime)]MTE(v^*) + \int_{p(z\prime)}^{p(z)}b|v-v^*|dv \\ 
  &\leq& [p(z)-p(z\prime)]MTE(v^*) + \int_{p_l(z\prime)}^{p_u(z)}b|v-v^*|dv 
\end{eqnarray*}
\normalsize
Symmetrically, 
\begin{eqnarray*}
  E[Y|Z=z]-E[Y|Z=z\prime] &\geq& 
   [p(z)-p(z\prime)]MTE(v^*) - \int_{p_l(z\prime)}^{p_u(z)}b|v-v^*|dv
\end{eqnarray*}
Then by a similar display as in the discussion before the theorem of the text we can get bounds based on no information about the sign, or based on the information contained in $E[Y|Z=z]-E[Y|Z=z\prime] \pm \int_{p_l(z\prime)}^{p_u(z)}b|v-v^*|dv$. A similar logic applies for $p(z)< v^*$ and $v^*<p(z\prime)$. The following theorem summarizes this result. 

\footnotesize
\begin{theorem}\label{TLM}
If assumptions 2.1-2.2 and 2.4 holds. Then the following bounds are valid: 
\begin{enumerate}
   
\item If $E[Y|Z=z]-E[Y|Z=z\prime]-  \int_{p_l(z\prime)}^{p_u(z)}b|v-v^*|dv\geq 0$:
\begin{eqnarray*}
MTE^+(v^*)_{lb}=\frac{E[Y|Z=z]-E[Y|Z=z\prime]-  \int_{p_l(z\prime)}^{p_u(z)}b|v-v^*|dv}{\Delta_{pu}} \\
MTE^+(v^*)_{ub}=\frac{E[Y|Z=z]-E[Y|Z=z\prime]+  \int_{p_l(z\prime)}^{p_u(z)}b|v-v^*|dv}{\Delta_{pl}}
\end{eqnarray*}
\item If $E[Y|Z=z]-E[Y|Z=z\prime]+  \int_{p_l(z\prime)}^{p_u(z)}b|v-v^*|dv\leq 0$:
\begin{eqnarray*}
MTE^-(v^*)_{lb}=\frac{E[Y|Z=z]-E[Y|Z=z\prime]-  \int_{p_l(z\prime)}^{p_u(z)}b|v-v^*|dv}{\Delta_{pl}} \\
MTE^-(v^*)_{ub}=\frac{E[Y|Z=z]-E[Y|Z=z\prime]+  \int_{p_l(z\prime)}^{p_u(z)}b|v-v^*|dv}{\Delta_{pu}}
\end{eqnarray*}
\item Otherwise: 
\begin{eqnarray*}
MTE(v^*)_{lb}=\max\{MTE^-(v^*)_{lb},MTE^+(v^*)_{lb}\} \\
MTE(v^*)_{ub}=\min\{MTE^-(v^*)_{ub},MTE^+(v^*)_{ub}\}
\end{eqnarray*}
\end{enumerate}

\end{theorem}
\normalsize

 \section{Identification with instruments taking more than 2 values}\label{appDis}
Consider the case where instead of the instrument taking either the value $0$ or $1$ now it takes values $\mathcal{Z}=\{z_1,z_2.....,z_K\}$. The developed identification strategy can be extended in this case. We will still be using the identified quantities $E[Y|Z=z]-E[Y|Z=z\prime]$ as the main input for identification. Given we now have $K$ possible values and we take combinations $z_i,z_j$ to obtain $E[Y|Z=z]-E[Y|Z=z\prime]$, we will have $\frac{K^2-K}{2}$ elements of the form   $E[Y|Z=z]-E[Y|Z=z\prime]$. 
\par
Define: 
\begin{eqnarray*}
\Delta_{1y}&=&E[Y|Z=z_1]-E[Y|Z=z_2] \\
\Delta_{2y}&=&E[Y|Z=z_1]-E[Y|Z=z_3] \\
&\vdots& \\
\Delta_{K-1y}&=&E[Y|Z=z_1]-E[Y|Z=z_K] \\
\Delta_{Ky}&=&E[Y|Z=z_2]-E[Y|Z=z_3] \\
&\vdots& \\
\Delta_{\frac{K^2-K}{2}y}&=&E[Y|Z=z_{K-1}]-E[Y|Z=z_K] 
\end{eqnarray*}
Similarly, define $\Delta_{1p}=p(z_1)-p(z_2), \Delta_{1v^*}=2b\int_{p_l(z_2)}^{p_u(z_1)}|v-v^*|dv$. Furthermore let $\Delta_{1up}$ and $\Delta_{1lp}$ be respectively the upper and lower bound of the difference between $p(z_1), p(z_2)$. Define similarly the quantities for the other combinations of instruments. Following the display of the main text we then have the following set of equations for any given $v^*$ of interest:  

\begin{eqnarray}
\Delta_{1y}- \Delta_{1v^*} &\leq& MTE(v^*)\Delta_{1p}  \label{S0} \\
&\vdots&  \nonumber \\
\Delta_{\frac{K^2-K}{2}y}- \Delta_{\frac{K^2-K}{2}v^*} &\leq & MTE(v^*)\Delta_{\frac{K^2-K}{2}p} \nonumber \\
\Delta_{1y}+ \Delta_{1v^*} &\geq& MTE(v^*)\Delta_{1p} \nonumber \\
&\vdots&  \nonumber \\
\Delta_{\frac{K^2-K}{2}y}+ \Delta_{\frac{K^2-K}{2}v^*} &\geq & MTE(v^*)\Delta_{\frac{K^2-K}{2}p}  \nonumber
\end{eqnarray}
As before the previous equations contain sufficient conditions for the sign of the $MTE$ at that particular $v^*$. These can be exploited in the following way: 
\footnotesize
\begin{eqnarray}
\textrm{If} \max_k \{\Delta_{1y}- \Delta_{1v^*},....,\Delta_{ky}- \Delta_{kv^*},.....,\Delta_{\frac{K^2-K}{2}y}-\Delta_{\frac{K^2-K}{2}v^*}\} \geq 0 \quad \
\textrm{then} \quad MTE(v^*) \geq 0 \quad \label{S1}  \\ 
\textrm{If} \min_k \{\Delta_{1y}+ \Delta_{1v^*},....,\Delta_{ky}+ \Delta_{kv^*},.....,\Delta_{\frac{K^2-K}{2}y}+\Delta_{\frac{K^2-K}{2}v^*}\}\leq 0 \quad
\textrm{then} \quad MTE(v^*) \leq 0 \quad \label{S2}
\end{eqnarray}
\normalsize
Solving  the system \ref{S0} for $MTE(v^*)$ and taking in consideration \ref{S1}-\ref{S2} we know the following bounds summarized in this proposition: 
\begin{proposition}\label{Prop1}
If assumptions 2.1-2.2 and 2.4 holds. Then the following bounds are valid: 
\begin{enumerate}
    \item If \ref{S1} holds, then: 
    \begin{eqnarray*}
MTE^+(v^*)_{lb}&=&\max\{\frac{\Delta_{1y}- \Delta_{1v^*}}{\Delta_{1up}},......, \frac{\Delta_{\frac{K^2-K}{2}y}- \Delta_{\frac{K^2-K}{2}v^*}}{\Delta_{\frac{K^2-K}{2}up}} \} \\
MTE^+(v^*)_{ub}&=&\min\{ \frac{\Delta_{1y}+ \Delta_{1v^*}}{\Delta_{1lp}},......, \frac{\Delta_{\frac{K^2-K}{2}y}+ \Delta_{\frac{K^2-K}{2}v^*}}{\Delta_{\frac{K^2-K}{2}lp}} \}
\end{eqnarray*}
\item If \ref{S2} holds, then: 
    \begin{eqnarray*}
MTE^-(v^*)_{lb}&=&\max\{\frac{\Delta_{1y}- \Delta_{1v^*}}{\Delta_{1lp}},......, \frac{\Delta_{\frac{K^2-K}{2}y}- \Delta_{\frac{K^2-K}{2}v^*}}{\Delta_{\frac{K^2-K}{2}lp}} \} \\
MTE^-(v^*)_{ub}&=&\min\{ \frac{\Delta_{1y}+ \Delta_{1v^*}}{\Delta_{1up}},......, \frac{\Delta_{\frac{K^2-K}{2}y}+ \Delta_{\frac{K^2-K}{2}v^*}}{\Delta_{\frac{K^2-K}{2}up}} \}
\end{eqnarray*}
\item Otherwise: 
\begin{eqnarray*}
MTE(v^*)_{lb}=\max\{MTE^-(v^*)_{lb},MTE^+(v^*)_{lb}\} \\
MTE(v^*)_{ub}=\min\{MTE^-(v^*)_{ub},MTE^+(v^*)_{ub}\}
\end{eqnarray*}
\end{enumerate}
\end{proposition}
\subsection*{The special case of the instrument taking 3 values and the treatment cannot hurt assumption}
\noindent In this subsection the previous bounds are computed for the particular  case that $\mathcal{Z}=\{ z_0,z_1,z_2\}$, $P(Y_1<Y_0)=1$, the upper bound of $Y$ is $1$, the lower bound is $0$ and $p_{l0} \leq p_{l1} \leq  p_{u1} \leq  p_{u2}$. This computation pretends to illustrate how the bounds would look like in this particular case and also serves the empirical application. In such application we impose the treatment cannot hurt assumption ($P(Y_1<Y_0)=1$) and the point estimates for $\alpha=10\%$ of the upper and lower bounds of the propensity scores are consistent with  $p_{l0} \leq p_{l1} \leq  p_{u1} \leq  p_{u2}$. Also the outcome $Y$ in the application has a bounded support. Computing the bounds under these conditions we get the following result for the different positions of the $v^*$ of interest:
\begin{enumerate}
    \item If $v^* \leq p_{l0} \leq p_{l1} \leq  p_{u1} \leq  p_{u2}$. Then: 
       \begin{eqnarray*}
MTE^-(v^*)_{lb}&=&\max \Bigg\{-1, \frac{\Delta_{yz_2z_1}-2b[\frac{p_{u2}^2-p_{l1}^2}{2}+v^*(p_{l1}-p_{u2})]}{\Delta_{lpz_2z_1}}, \\& &  \frac{\Delta_{yz_2z_0}-2b[\frac{p_{u2}^2-p_{l0}^2}{2}+v^*(p_{l0}-p_{u2})]}{\Delta_{lpz_2z_0}}, \frac{\Delta_{yz_1z_0}-2b[\frac{p_{u1}^2-p_{l0}^2}{2}+v^*(p_{l0}-p_{u1})]}{\Delta_{lpz_1z_0}}  \Bigg\} \\
MTE^-(v^*)_{ub}&=&\min\Bigg\{0, \frac{\Delta_{yz_2z_1}+2b[\frac{p_{u2}^2-p_{l1}^2}{2}+v^*(p_{l1}-p_{u2})]}{\Delta_{upz_2z_1}}, \\& & \frac{\Delta_{yz_2z_0}+2b[\frac{p_{u2}^2-p_{l0}^2}{2}+v^*(p_{l0}-p_{u2})]}{\Delta_{upz_2z_0}}, \frac{\Delta_{yz_1z_0}+2b[\frac{p_{u1}^2-p_{l0}^2}{2}+v^*(p_{l0}-p_{u1})]}{\Delta_{upz_1z_0}}  \Bigg\}
\end{eqnarray*}
    \item If  $ p_{l0} \leq v^* \leq p_{l1} \leq  p_{u1} \leq  p_{u2}$. Then: 
       \begin{eqnarray*}
MTE^-(v^*)_{lb}&=&\max\Bigg\{-1, \frac{\Delta_{yz_2z_1}-2b[\frac{p_{u2}^2-p_{l1}^2}{2}+v^*(p_{l1}-p_{u2})]}{\Delta_{lpz_2z_1}}, \\& &  \frac{\Delta_{yz_2z_0}-2b[v^{*2}-v^*(p_{u2}+p_{l0})+\frac{p_{u2}^2+p_{l0}^2}{2}]}{\Delta_{lpz_2z_0}}, \frac{\Delta_{yz_1z_0}-2b[v^{*2}-v^*(p_{u1}+p_{l0})+\frac{p_{u1}^2+p_{l0}^2}{2}]}{\Delta_{lpz_1z_0}} \Bigg \} \\
MTE^-(v^*)_{ub}&=&\min\Bigg\{0, \frac{\Delta_{yz_2z_1}+2b[\frac{p_{u2}^2-p_{l1}^2}{2}+v^*(p_{l1}-p_{u2})]}{\Delta_{upz_2z_1}}, \\& &  \frac{\Delta_{yz_2z_0}+2b[v^{*2}-v^*(p_{u2}+p_{l0})+\frac{p_{u2}^2+p_{l0}^2}{2}]}{\Delta_{upz_2z_0}}, \frac{\Delta_{yz_1z_0}+2b[v^{*2}-v^*(p_{u1}+p_{l0})+\frac{p_{u1}^2+p_{l0}^2}{2}]}{\Delta_{upz_1z_0}}  \Bigg \}
\end{eqnarray*}
 \item If  $ p_{l0}  \leq p_{l1} \leq  v^* \leq  p_{u1} \leq  p_{u2}$. Then: 
       \begin{eqnarray*}
MTE^-(v^*)_{lb}&=&\max\Bigg\{-1, \frac{\Delta_{yz_2z_1}-2b[v^{*2}-v^*(p_{u2}+p_{l1})+\frac{p_{u2}^2+p_{l1}^2}{2}]}{\Delta_{lpz_2z_1}}, \\& &  \frac{\Delta_{yz_2z_0}-2b[v^{*2}-v^*(p_{u2}+p_{l0})+\frac{p_{u2}^2+p_{l0}^2}{2}]}{\Delta_{lpz_2z_0}}, \frac{\Delta_{yz_1z_0}-2b[v^{*2}-v^*(p_{u1}+p_{l0})+\frac{p_{u1}^2+p_{l0}^2}{2}]}{\Delta_{lpz_1z_0}}  \Bigg\} \\
MTE^-(v^*)_{ub}&=&\min\Bigg\{0, \frac{\Delta_{yz_2z_1}+2b[v^{*2}-v^*(p_{u2}+p_{l1})+\frac{p_{u2}^2+p_{l1}^2}{2}]}{\Delta_{upz_2z_1}}, \\& &  \frac{\Delta_{yz_2z_0}+2b[v^{*2}-v^*(p_{u2}+p_{l0})+\frac{p_{u2}^2+p_{l0}^2}{2}]}{\Delta_{upz_2z_0}}, \frac{\Delta_{yz_1z_0}+2b[v^{*2}-v^*(p_{u1}+p_{l0})+\frac{p_{u1}^2+p_{l0}^2}{2}]}{\Delta_{upz_1z_0}}   \Bigg \}
\end{eqnarray*}
 \item If  $ p_{l0}  \leq p_{l1} \leq    p_{u1} \leq v^*  \leq  p_{u2}$. Then: 
       \begin{eqnarray*}
MTE^-(v^*)_{lb}&=&\max\Bigg\{-1, \frac{\Delta_{yz_2z_1}-2b[v^{*2}-v^*(p_{u2}+p_{l1})+\frac{p_{u2}^2+p_{l1}^2}{2}]}{\Delta_{lpz_2z_1}}, \\& &  \frac{\Delta_{yz_2z_0}-2b[v^{*2}-v^*(p_{u2}+p_{l0})+\frac{p_{u2}^2+p_{l0}^2}{2}]}{\Delta_{lpz_2z_0}}, \frac{\Delta_{yz_1z_0}-2b[\frac{p_{l0}^2-p_{u1}^2}{2}+v^*(p_{u1}-p_{l0})]}{\Delta_{lpz_1z_0}}  \Bigg\} \\
MTE^-(v^*)_{ub}&=&\min\Bigg\{0, \frac{\Delta_{yz_2z_1}+2b[v^{*2}-v^*(p_{u2}+p_{l1})+\frac{p_{u2}^2+p_{l1}^2}{2}]}{\Delta_{upz_2z_1}}, \\& &  \frac{\Delta_{yz_2z_0}+2b[v^{*2}-v^*(p_{u2}+p_{l0})+\frac{p_{u2}^2+p_{l0}^2}{2}]}{\Delta_{upz_2z_0}}, \frac{\Delta_{yz_1z_0}+2b[\frac{p_{l0}^2-p_{u1}^2}{2}+v^*(p_{u1}-p_{l0})]}{\Delta_{upz_1z_0}}   \Bigg\}
\end{eqnarray*}

 \item If  $ p_{l0}  \leq p_{l1} \leq    p_{u1}   \leq  p_{u2} \leq v^*$. Then: 
       \begin{eqnarray*}
MTE^-(v^*)_{lb}&=&\max\Bigg\{-1, \frac{\Delta_{yz_2z_1}-2b[\frac{p_{l1}^2-p_{u2}^2}{2}+v^*(p_{u2}-p_{l1})]}{\Delta_{lpz_2z_1}}, \\& &  \frac{\Delta_{yz_2z_0}-2b[\frac{p_{l0}^2-p_{u2}^2}{2}+v^*(p_{u2}-p_{l0})]}{\Delta_{lpz_2z_0}}, \frac{\Delta_{yz_1z_0}-2b[\frac{p_{l0}^2-p_{u1}^2}{2}+v^*(p_{u1}-p_{l0})]}{\Delta_{lpz_1z_0}}  \Bigg\} \\
MTE^-(v^*)_{ub}&=&\min\Bigg\{0, \frac{\Delta_{yz_2z_1}+2b[\frac{p_{l1}^2-p_{u2}^2}{2}+v^*(p_{u2}-p_{l1})]}{\Delta_{upz_2z_1}}, \\& &  \frac{\Delta_{yz_2z_0}+2b[\frac{p_{l0}^2-p_{u2}^2}{2}+v^*(p_{u2}-p_{l0})]}{\Delta_{upz_2z_0}}, \frac{\Delta_{yz_1z_0}+2b[\frac{p_{l0}^2-p_{u1}^2}{2}+v^*(p_{u1}-p_{l0})]}{\Delta_{upz_1z_0}}  \Bigg\}
\end{eqnarray*}
\begin{remark}
When the values the instrument take grows, it can be cumbersome to compute the integral, so instead of that, a researcher might be willing to solve it numerically and apply proposition \ref{Prop1}. 
\end{remark}
\end{enumerate}

 \section{Illustration of results in appendix \ref{appDis}}\label{appDis2}
\begin{figure}[ht]
\centering
 \caption{Bounds for the $MTE$ with a discrete $IV$, $b=0.2$ and $\alpha=0.1$ ($Y$ axis is the $MTE$ and $X$ axis is the value of $v^*$)}\label{F3}
\includegraphics[width=.9\textwidth]{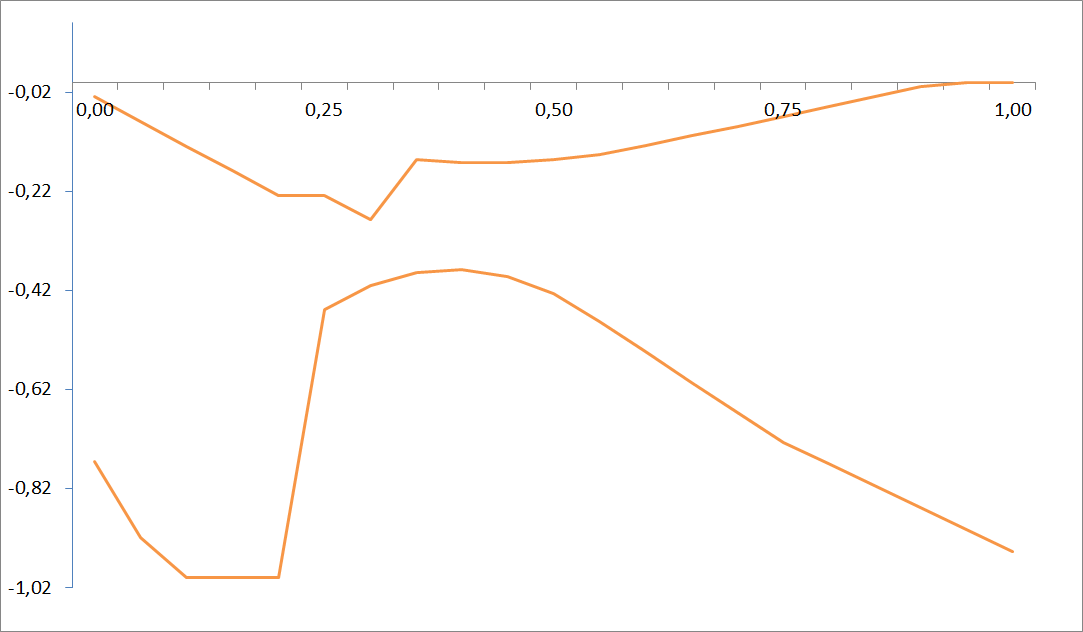}
\end{figure}

\begin{table}[ht]
\caption{Bounds on the $ATE$, comparison with extra values of the IV for $b=0.5, \alpha=0.1$}
\centering 
\begin{tabular}{|cc|cc|}
\hline
\multicolumn{2}{|c|}{\textit{3-valued   IV}}       & \multicolumn{2}{c|}{\textit{2-valued IV}}       \\ \hline
\multicolumn{1}{|c|}{\textit{Lower  Bound}}  & \textit{Upper Bound}  & \multicolumn{1}{c|}{\textit{Lower Bound}}  & \textit{Upper Bound} \\ \hline
\multicolumn{1}{|c|}{-0,7}          & -0,1         & \multicolumn{1}{c|}{-0,8}         & 0,0         \\ \hline
\multicolumn{2}{|c|}{\textit{Lenght   of the set}} & \multicolumn{2}{c|}{\textit{Lenght of the set}} \\ \hline
\multicolumn{2}{|c|}{0,6}                          & \multicolumn{2}{c|}{0,8}                        \\ \hline
\end{tabular}
\end{table}
See Figure \ref{F3}
\section{Figures}\label{appfig}
\begin{figure}[ht]
\centering
 \caption{Bounds for the $MTE$ ($Y$ axis is the $MTE$ and $X$ axis is the value of $v^*$)} \label{F1}

\includegraphics[width=.5\textwidth]{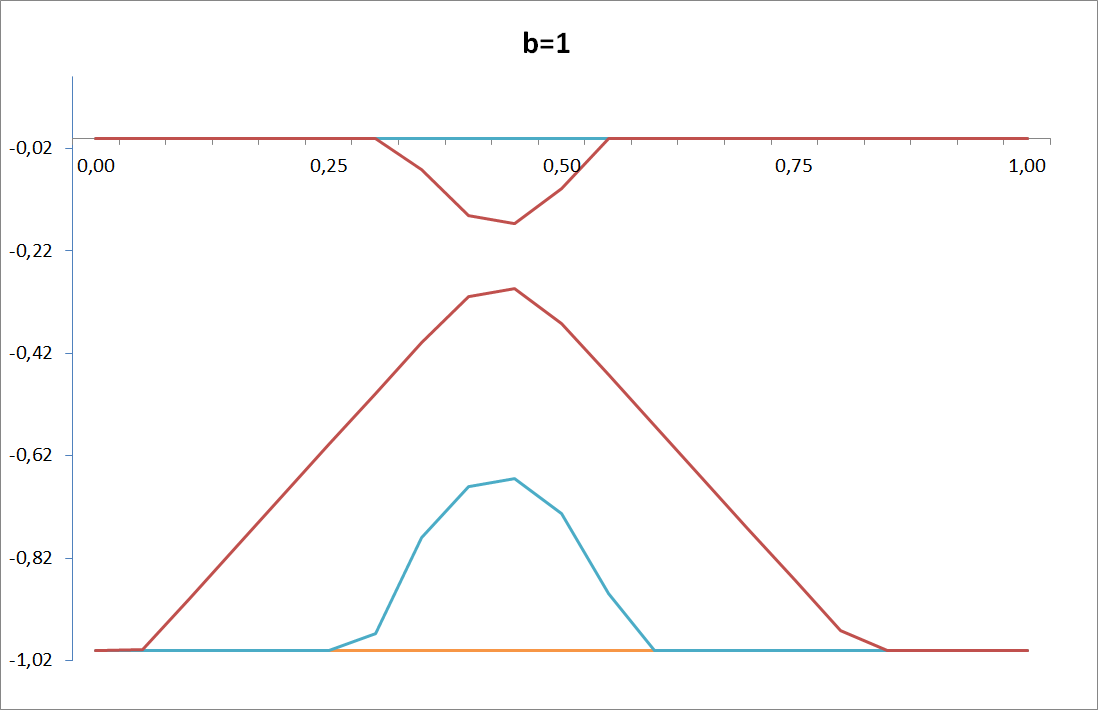}\hfill

\includegraphics[width=.5\textwidth]{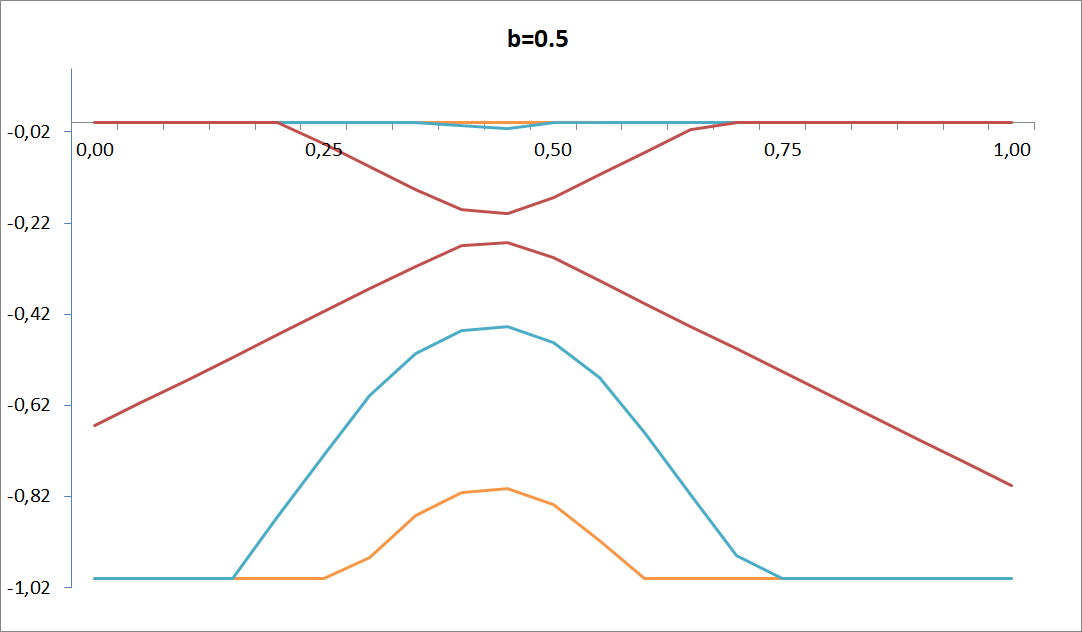}\hfill
\includegraphics[width=.5\textwidth]{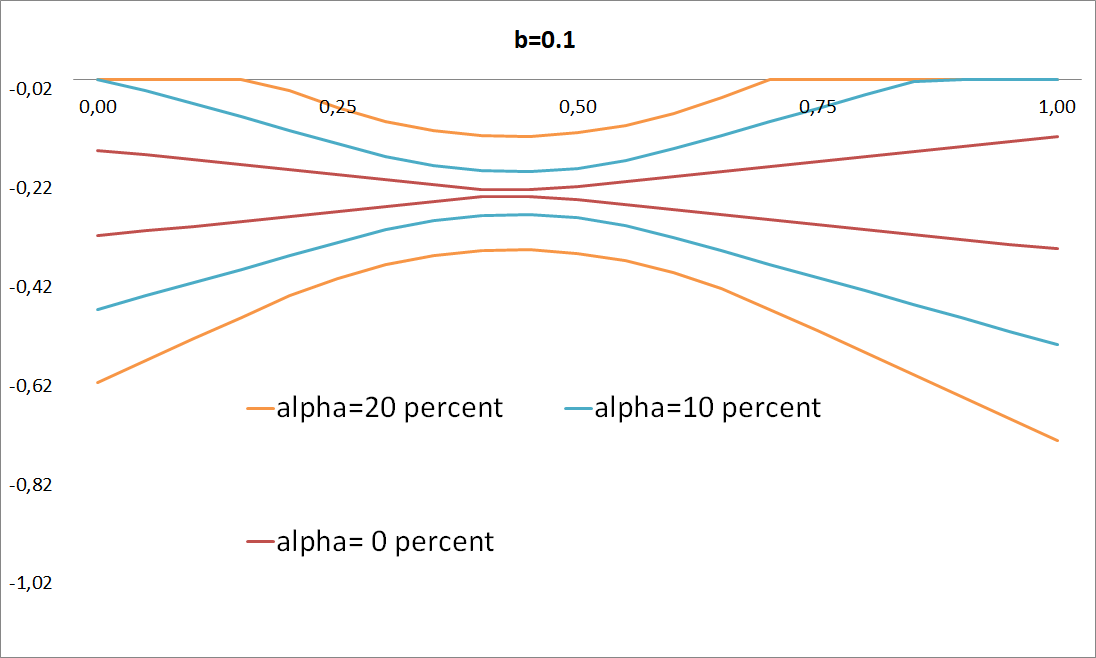}
  
\end{figure}


\end{document}